\documentclass[aps,twocolumn,prb,superscriptaddress,floatfix]{revtex4-1}
\usepackage{amssymb,amsmath}
\usepackage{graphicx}
\usepackage{dcolumn}
\usepackage{float}
\usepackage{bm}
\usepackage{pifont}
\usepackage{cancel}
\usepackage{latexsym}
\usepackage{textcomp}
\usepackage{array}
\usepackage{amssymb}
\usepackage{amsfonts}
\usepackage{dsfont}
\usepackage{colortbl}
\usepackage{afterpage}
\usepackage{amsmath}
\usepackage{fancybox}
\usepackage{color}
\usepackage{afterpage}
\usepackage[vcentermath, noautoscale]{youngtab}
\usepackage{ulem}
\usepackage{textcase}
\usepackage[printonlyused]{acronym}
\definecolor{MyDarkBlue}{rgb}{0,0.08,0.45}
\definecolor{darkgreen}{rgb}{0,0.4,0.00}

\newcolumntype{X}{>{\setbox0=\hbox\bgroup}c<{\egroup}@{}}

\newcommand{\tofinish}[1]{}
\newcommand{\missing}[1]{}
\newcommand{\inspicture}[1]{}
\newcommand{\ask}[1]{}
\newcommand{\important}[1]{}
\newcommand{\correction}[1]{}
\newcommand{\suggestion}[1]{}

\newcommand{\sqbra}[1]{\left[#1\right]}

\newcommand{\up}{\uparrow}
\newcommand{\dw}{\downarrow}
\newcommand{\de}[1]{d{#1}\,}
\newcommand{\Tr}[1]{\operatorname{Tr}\left\{#1\right\}}
\newcommand{\ocd}{\hat{c}^{\dagger}}

\renewcommand{\oc}{\hat{c}^{\phantom{\dagger}}}

\newcommand{\dagga}{{\phantom{\dagger}}}
\newcommand{\be}{\begin{equation}}
\newcommand{\ee}{\end{equation}}
\newcommand{\bea}{\begin{eqnarray}}
\newcommand{\eea}{\end{eqnarray}}
\newcommand{\baa}{\begin{align}}
\newcommand{\eaa}{\end{align}}
\newcommand{\br}{{\bm r}}

\newcommand{\brp}{{\bm r'}}
\newcommand{\bR}{{\bm R}}
\newcommand{\bRp}{{\bm R'}}
\newcommand{\bk}{{\bm k}}

\newcommand{\fract}[2]{\displaystyle{\frac{#1}{#2}}}
\newcommand{\eqn}[1]{Eq.~\eqref{#1}}
\newcommand{\fig}[1]{Fig.~\ref{#1}}

\newcommand{\rtab}[1]{Table~\ref{#1}}

\newcommand{\sect}[1]{Sect.~\ref{#1}}
\newcommand{\appendixx}[1]{Appendix~\ref{#1}}

\newcommand{\expect}[2]{\langle {#1|#2|#1}\rangle}

\newcommand{\ket}[1]{|#1\,\rangle}
\newcommand{\bra}[1]{\langle\,#1|}

\newcommand{\HKfunc}{F_{\rm HK}}

\newcommand{\kPsiG}{\ket{\Psi_{\rm G}}}
\newcommand{\kPsio}{\ket{\Psi_{0}}}

\newcommand{\Psio}{\Psi_{0}}
\newcommand{\PsiG}{{\Psi_{\rm G}}}
\newcommand{\Ham}{\hat{H}}
\newcommand{\hHub}{\hat{H}_{\rm Hub}}

\newcommand{\hHund}{\hat{H}_{\rm Hund}}
\newcommand{\hHat}{\hat{H}_{\rm at}}
\newcommand{\hHint}{\hat{H}_{\rm int}}

\newcommand{\Phig}{{\Phi}}

\newcommand{\hPhig}{\hat{\Phi}^{\phantom{\dagger}}}

\newcommand{\hPhigd}{\hat{\Phi}^\dagger}

\newcommand{\ra}{\rightarrow}

\newcommand{\Gpg}[1]{\hat{\cal P}_{#1}}
\newcommand{\Gpl}[1]{\hat{\cal P}_{#1}}

\newcommand{\balpha}{{\bm{\alpha}}}
\newcommand{\bbeta}{{\bm{\beta}}}

\newcommand{\hN}{\hat{N}}
\newcommand{\Nup}{N_{\up}}

\newcommand{\Ndw}{N_{\dw}}

\newcommand{\ndw}{{n_{\dw}}}
\newcommand{\nup}{{n_{\up}}}

\newcommand{\hident}{\hat{\mathds{1}}}

\newcommand{\hn}{{\hat{n}}}

\newcommand{\Numbei}{N}
\newcommand{\Elz}{\hat{L}_{z}}
\newcommand{\Elsq}{\hat{L}^2}
\newcommand{\telsq}{L}

\newcommand{\telei}{L(L+1)}
\newcommand{\Elpl}{\hat{L}_{+}}
\newcommand{\Elmi}{\hat{L}_{-}}

\newcommand{\Spinz}{{\hat{S}_{z}}}

\newcommand{\Spinsq}{S^2}

\newcommand{\tspinei}{S(S+1)}
\newcommand{\tspinsq}{S}

\newcommand{\tspinz}{S_z}
\newcommand{\eg}{e_{g}}
\newcommand{\tdg}{t_{2g}}

\newcommand{\Sudu}{$SU(2)$}
\newcommand{\Otre}{$O(3)$}

\newcommand{\Lam}{\Lambda}

\newcommand{\EGS}{E_{\rm GS}}

\newcommand{\Do}{D^{(0)}}
\newcommand{\D}{{D}}
\newcommand{\hT}{\hat{T}}

\newcommand{\no}{n^{(0)}}
\newcommand{\nnat}{n^{(0)}}
\newcommand{\hVee}{\hat{V}_{\rm ee}}
\newcommand{\hVext}{\hat{V}_{\rm ext}}
\newcommand{\Vext}{V_{\rm ext}}

\newcommand{\Vextm}{V^{(\rm ext)}}
\newcommand{\Ts}{T_{s}}

\newcommand{\EH}{E_{\rm H}}

\newcommand{\DF}{{\cal F}}

\newcommand{\Exc}{E_{\rm xc}}

\newcommand{\EHub}{E_{\rm Hub}}
\newcommand{\Eat}{E_{\rm at}}
\newcommand{\Ekin}{E_{\rm kin}}

\newcommand{\Exco}{E^{(0)}_{\rm xc}}
\newcommand{\EHo}{E^{(0)}_{\rm H}}

\newcommand{\Edc}{{E_{\rm dc}}}

\newcommand{\vxc}{v_{\rm xc}}

\newcommand{\Vo}{V^{(\rm 0)}}
\newcommand{\vH}{v_{\rm H}}
\newcommand{\VHm}{V^{(\rm H)}}

\newcommand{\Vxcm}{V^{(\rm xc)}}

\newcommand{\Gname}{Martin C. Gutzwiller}

\newcommand{\xGm}{Gutzwiller method}

\newcommand{\irrep}{irreducible representation}
\newcommand{\xGp}{Gutzwiller parameter matrix}

\newcommand{\xGps}{Gutzwiller parameters}
\newcommand{\xno}{natural density matrix}

\newcommand{\acwk}{and coworkers}

\usepackage{ifthen}
\usepackage{calc}
\usepackage{url}
\usepackage{fancyvrb}
\usepackage[utf8]{inputenc}
\usepackage[english]{babel}%

\newacro{ldag}[LDA+G]{Local Density Approximation plus Gutzwiller Method}
\newacro{lda}[LDA]{Local Density Approximation}
\newacro{dmft}[DMFT]{Dynamical Mean-Field Theory}
\newacro{ldau}[LDA+U]{Local Density Approximation plus Hubbard-$U$}
\newacro{ga}[GA]{Gutzwiller Approximation}
\newacro{dft}[DFT]{Density Functional Theory}
\newacro{gvm}[GVM]{Gutzwiller Variational Method}
\newacro{ldadmft}[LDA+DMFT]{Local Density Approximation plus Dynamical Mean-Field Theory}
\newacro{lsda}[LSDA]{Local Spin Density Approximation}
\newacro{gga}[GGA]{Generalized Gradient Approximation}
\newacro{hm}[HM]{Hubbard model}
\newacro{ga}[GA]{Gutzwiller Approximation}
\newacro{lm}[LM]{Levenberg-Marquardt}
\newacro{hf}[HF]{Hartree-Fock}
\newacro{vqmc}[VQMC]{Variational Quantum Monte Carlo}
\newacro{si}[SI]{self-interaction}
\newacro{bec}[BC]{basis of electronic configurations}
\newacro{mbsb}[MBSB]{many-body symmetric basis}
\begin{document}
\title{Self-consistent Gutzwiller study of $bcc$ Fe: interplay of ferromagnetic order and kinetic energy }
\author{Giovanni Borghi}\affiliation{Theory and Simulation of Materials, \'Ecole Polytechnique F\'ed\'erale de Lausanne EPFL, CH-1015 Lausanne, Switzerland.}
\affiliation{International School for advanced studies (SISSA), via Bonomea 265, 34136 Trieste, Italy.} 
\author{Michele Fabrizio}\affiliation{International School for advanced studies (SISSA), via Bonomea 265, 34136 Trieste, Italy.}
\author{Erio Tosatti}
\affiliation{International School for advanced studies (SISSA), via Bonomea 265, 34136 Trieste, Italy.}
\affiliation{CNR-IOM Democritos, via Bonomea 265, 34136 Trieste, Italy.}
\affiliation{International Centre for Advanced Studies, Strada Costiera 11, 34151 Trieste, Italy.}
\email{giovanni.borghi@epfl.ch, fabrizio@sissa.it, tosatti@sissa.it}
\begin{abstract}

The Gutzwiller technique has long been known as a method to include correlations in electronic structure calculations. 
Here we implement an {\sl ab-initio} Gutzwiller+LDA calculation exposing the detailed protocol step by step, in a way 
that should be of use for future applications. 
We apply it to a classic problem, the ferromagnetism of bulk $bcc$ Fe, whose nature has attracted recent interest.
In the conventional Stoner-Wohlfarth model, the ferromagnetic ordering of iron sets in so that the electrons can reduce 
their mutual Coulomb repulsion-- naturally at the extra cost of some increase of electron kinetic energy.  Density functional 
theory within the spin polarized local density approximation (LDA) calculations has long supported that picture, showing 
that ferromagnetic alignment causes band narrowing and a corresponding wavefunction localization, 
whence a kinetic energy increase. 
However, because of its inadequate treatment of strong intra-site  correlations for localized $d$ orbitals, LDA cannot 
be relied upon, particularly when it comes to separately describing fine potential 
and kinetic energy imbalances.  With {\sl ab-initio} Gutzwiller+LDA, we indeed find that the effect of correlations 
is to dramatically reverse the balance, the ferromagnetic ordering of Fe in fact causing a decrease of kinetic energy, 
at the cost of some increase of potential energy. The underlying physical mechanism, foreshadowed long ago by 
Goodenough and others, and more recently supported by LDA-DMFT calculations, is that correlations cause $e_g$ 
and $t_{2g}$ $3d$ orbitals  to behave very differently. Weakly dispersive $e_g$ states are spin-polarized and almost 
localized, while, more than half filled, the $t_{2g}$ are broad band, fully delocalized states. Owing to intra-atomic 
Hund's rule exchange which aligns $e_g$ and $t_{2g}$ spins, the propagation of itinerant $t_{2g}$ holes is only allowed 
when different atomic spins are ferromagnetically aligned. We thus conclude that double exchange is at work already in the 
most popular ferromagnetic metal. 

\end{abstract}
\maketitle

\section*{Introduction}\label{Sec:intro}

The conduction electron Wannier orbitals in transition-metal compounds are generally  
fairly localized in space so that electronic correlations, i.e. all effects that deviate from the 
independent-particle picture,  are sometimes strong enough  to give rise to metal-insulator 
transitions in particular temperature and pressure conditions. The correlation-driven 
metal-insulator transition, known as Mott transition,\cite{Mott_original, Mott} is 
often accompanied by rather spectacular phenomena that appear in its proximity, high-temperature 
superconductivity being the most popular example. This makes $3d$ metal 
elements and 
compounds a natural laboratory for intriguing many-body physics, which despite 
a rich history and many studies is worth exploring further. 

Electronic structure methods (sometimes referred to as "first principles" methods) that rely on independent 
particle descriptions, such as \ac{hf} or \ac{dft} within \ac{lda}, are by construction 
incapable of capturing the Mott transition, 
which has no counterpart in a one electron picture.
For this reason, \ac{hf} and \ac{lda},  while 
generally quite successful for many materials,
may sometimes fail in the description of solids involving transition metals.  In fact, most of our knowledge 
about Mott electron localization has been attained by means of simplified lattice models, the best known 
being the Hubbard model,\cite{Hubbard} which are accessible by methods better suited to deal with correlations, 
such as quantum Monte Carlo,\cite{QMC_review} density-matrix renormalization group\cite{Schwollock_DMRG} 
and dynamical mean-field theory.\cite{DMFT} 

Clearly, for  the purpose of a quantitative understanding of real materials, it is of key importance to 
sew the two worlds together, bringing in particular the many body
expertise gained on lattice models over to realistic, off-lattice, first-principles calculations of solids.

This has historically been attempted through ad-hoc improvements of \ac{dft}. 
For instance, the inclusion (in fact, the addition and subtraction) of an intra-site Coulomb  repulsion $U$ (the "Hubbard $U$") 
in the Kohn-Sham Hamiltonian permits a decrease of the so-called self-interaction error,  
a severe flaw of \ac{lda} for partially or fully occupied  localized orbitals -- just the case of transition metals.  
When added to \ac{lda}, this procedure, the so called 
\ac{ldau},\cite{LDAU_Anisimov_Andersen, LDAU_jphys} often improves results, and can for example stabilize magnetic phases which 
straight \ac{lda} would miss. Yet, \ac{ldau} remains basically a mean-field, independent particle approach that cannot 
describe  Mott localization. The problem can be overcome if, for instance, the Kohn-Sham Hamiltonian of \ac{lda} 
supplemented by $U$ is solved through \ac{dmft},  by the so-called \ac{ldadmft}.\cite{LDADMFT_theory} 
Alternatively, variational Quantum Monte Carlo \ac{vqmc} approaches\cite{Ceperley_VMC,Sorella_SR} have been successfully 
applied to  the electronic 
properties of atoms and simple molecules,\cite{Sorella_dimers} and its development appears to be promising 
for more ambitious applications.

At present, both \ac{ldadmft} and \ac{vqmc} are numerically much more involved and far more demanding 
than conventional \ac{lda} or even \ac{ldau}, which owe much of their success  
to simplicity. The desirability of approaches joining together the simplicity of \ac{lda} and the description of
correlations typical of many body methods is therefore still very high. 
In the context of lattice models, a simple approach to 
strong correlations was proposed long ago by \Gname.\cite{Gutzwiller1,Gutzwiller2} This method,
projecting out of a trial Slater determinant an adjustable proportion of costly configurations and 
evaluating average values by approximate formulas, is strictly variational in the limit of 
infinite lattice-coordination\cite{Gebhard} -- the same limit where \ac{dmft} is exact -- providing much more accurate 
results than \ac{hf}.  That success invites the use of the Gutzwiller method even when 
the lattice space dimension, and thus the site coordination, is finite, as people do with \ac{dmft}. 
\ac{ga}  electronic structure calculations 
have the great advantage 
to couple extreme LDA-level simplicity with qualitatively, often quantitatively, increased accuracy in the
description of correlations. For example, \ac{ga} has been able to describe conducting materials that are 
insulators ``in disguise'',\cite{Fazekas} i.e. whose properties depend on correlations that are already present 
in their Mott insulating phase, and that continue to play an important role even in the nearby metallic phases. 
A well known example is the RVB scenario for high-temperature superconductors,\cite{Anderson_RVB_Science} 
where Cooper pairing is explained as a byproduct of doping a parent state of resonating valence bonds, 
which is the remnant of antiferromagnetism when N\`eel long range order disappears. Another famous 
result of the \ac{ga} is the Brinkman-Rice description of the Mott transition in vanadium sesquioxide, originally 
derived by the \ac{ga} solution to the Hubbard model.\cite{brinkman&rice}. 

Because of its simplicity, a great deal of effort has therefore been 
devoted in recent years to extend \ac{ga} from simple lattice models to more realistic off-lattice 
cases.\cite{Ho_LDAG_condmat, ZhongFang_LDAG, Andersen_Gebhard_Gutzwiller, ZhongFang_LDAG_app1, Lanata_efficient, Ho,Lanata-Kotliar} 
Here we  implement  a density self-consistent algorithm that exploits the Gutzwiller variational wave 
function together with the conventional \ac{lda} for the density functional. The Levy-Lieb constrained-search 
formulation of \ac{dft} provides a solid theoretical framework for the introduction of Gutzwiller variational 
parameters in the density functional, while a localized atomic basis set (we use in particular the Siesta electronic structure code) makes 
the definition of the Gutzwiller-projected states straightforward.

We test the power of the \ac{ldag} functional by calculating the electronic structure of nonmagnetic
and ferromagnetic $bcc$ Fe, motivated  by long standing basic 
questions about the electronic origin of magnetic order,  
\cite{vollhardt2001} including 
a recent \ac{ldadmft} study 
by Anisimov and coworkers\cite{anisimov} 
suggesting that $bcc$ iron might be an orbital-selective Mott insulator. According to that picture,  the 
poorly dispersive $\eg$-type electrons of metallic Fe may be fully localized due to interactions, 
so that conduction phenomena are restricted within the $\tdg$ manifold (besides of course 
the $s$ electrons). In that picture\cite{stollhoff_ironloc, Goodenough_ironloc, stearns_ironloc} 
ferromagnetic alignment 
would not be due to inter-site Coulomb exchange, as is 
ordinarily assumed, but rather to double-exchange, as in colossal magnetoresistance manganites.\cite{manganites}
The Mott localized $e_g$ electrons 
form spin-1 moments that  couple 
ferromagnetically via intra-atomic Hund's exchange to the electrons in the  
nearly-full itinerant $t_{2g}$ bands. In order to preserve coherent $t_{2g}$ hole motion, 
the local $e_g$ moments order ferromagnetically. As in the manganites, ferromagnetism is 
thus driven by a kinetic energy gain rather than a potential energy one.   
Even though our \ac{ldag} approach is still mean-field and thus cannot address dynamical  
phenomena such as orbital selective Mott transitions --
especially so in a delicate case where the two sets of orbitals, $\eg$ and $\tdg$,  hybridize with each other 
in the Brillouin zone --  we find that calculation of the total energy and a detailed analysis of its 
separate kinetic and potential energy contributions actually supports  double-exchange 
as the driving mechanism of  
ferromagnetism in iron, rather than the conventional Stoner instability. On the whole, this work 
may also be of general use as a very detailed %represents a useful 
example of {\sl ab-initio} application of Gutzwiller correlations to a realistic electronic structure problem.

The plan of this article is as follows: in \sect{Sect1} we introduce the formalism of \ac{ldag} starting 
from the constrained-search formulation of Density Functional Theory, %and 
demonstrating
how the Gutzwiller 
wavefunction can be used to generalize \ac{ldau} by allowing the expectation value of the atomic 
Hamiltonian to be computed on a multi-determinant wavefunction. In \sect{Sect:Gutz_expect} and
\sect{Sec:Gutz_practice} we then show how the different terms of the \ac{ldag} density functional can be computed 
by means of \ac{ga}, and how the total energy of a correlated electronic system can be minimized by a three-step 
iterative procedure. In \sect{Sec:results} we finally present and comment on the physical results for paramagnetic and 
ferromagnetic $bcc$ Fe, and connect back to the basics questions about the origin of ferromagnetic order.

\section{Constrained-search formulation of a Gutzwiller Density Functional Theory}\label{Sect1}

A convenient way to introduce a Gutzwiller density functional is through the formalism 
independently proposed by Levy~\cite{Levy1,Levy2} and Lieb~\cite{Lieb1}.
Starting from the Rayleigh-Ritz definition for the ground state energy $\EGS$ of a system
\begin{align}\label{Eq_min_GSenergy}
\EGS = \min_{\Psi} \expect{\Psi}{\Ham}\,,
\end{align}
where the electron Hamiltonian $\Ham$ includes the kinetic energy $\hT$, the electron-electron interaction $\hVee$, 
and a local external potential $\hVext$, Levy and Lieb converted the variational principle for 
the ground state wavefunction into a variational principle for the ground state density through a constrained minimization at fixed density $n(\br)$
\begin{eqnarray}
\label{Eq_LL_DF}
\EGS[\Vext(\br)] &=& \min_{n(\br)} \Bigg\{ \min_{\Psi\ra n(\br)} \expect{\Psi}{\hT+\hVee} \nonumber \\
&& ~~~~~~~~~~+ \int \Vext(\br) n(\br)\Bigg\}\,.
\end{eqnarray}
The first term on the right-hand side of (\ref{Eq_LL_DF}) is nothing but the constrained-search definition of 
the Hohenberg-Kohn functional\cite{DFT_HKtheorem}, i.e.
\begin{align}\label{Eq_LL_HKfunc_def}
\HKfunc[n(\br)] = \min_{\Psi\ra n(\br)} \expect{\Psi}{\hT+\hVee}\,,
\end{align}
which is independent of the external potential $\Vext$.
The wavefunction $\Psi$ in the definition~\eqn{Eq_LL_HKfunc_def} should span the whole many-body Hilbert space, 
generally too large 
to allow a 
straightforward numerical evaluation of $\HKfunc[n(\br)]$.
Within the Kohn-Sham scheme, the generality of \eqn{Eq_LL_HKfunc_def} is abandoned in 
favor of a more practical definition of the Hohenberg and Kohn functional, in which the latter is split 
into kinetic, Hartree, and exchange-correlation terms, namely
\begin{align}\label{Eq_HK_semilocal}
\HKfunc[n(\br)] = \Ts[n(\br)] +\EH[n(\br)] + \Exc[n(\br)],
\end{align}
where $\EH[n(\br)]$ is simply the electrostatic energy of the electron density regarded as a classical charge distribution.
A constrained search is then retained only for the kinetic contribution
\begin{align}\label{Eq_HK_nonint}
\Ts[n(\br)] = \min_{\Psi\ra n(\br)} \expect{\Psi}{\hT}\,,
\end{align}
which, because $\hat{T}$ is a one-body operator, has a solution within the class of Slater determinants, 
a relatively simple task to accomplish through 
auxiliary non-interacting electron Hamiltonians whose ground state local density $n(\br)$ coincides 
with that of the physical interacting model. The insurmountable difficulties of the original many-body problem 
have thus been hidden in the unknown exchange-correlation functional $\Exc[n(\br)]$. All DFT approximation 
schemes correspond just to  different guesses of a physically sensible functional form of $\Exc[n(\br)]$ 
in terms of the local density. 

The main problem that arises from the density-dependent parametrization \eqn{Eq_HK_nonint} 
is that $\EH[n(\br)]$ contains a spurious \ac{si} 
term -- finite even when $n(\br)$ is the density of a single electron! -- a term which should be identically 
cancelled in the exact $\Exc[n(\br)]$. Unfortunately, all semi-local approximations to $\Exc[n(\br)]$, 
such as \ac{lda} and  \ac{gga}, fail to fully subtract such a \ac{si} term from the density functional, which 
brings about results that by construction contain a certain level of 
self-interaction 
error. 

The spurious SI one-electron energy is larger for spatially localized electronic wavefunctions. 
For instance, a single electron with a simple gaussian wavefunction feels an \ac{si} that is
inversely proportional to the standard deviation of the gaussian, only $70\%$ of which is 
subtracted by the \ac{lda} exchange functional. The improvements  attained by 
better functionals do not seem major.\cite{korzdorfer_SIC}
All density-functional calculations are affected to some extent by the \ac{si} error, more important 
when the real-space density matrix is more localized. That is especially 
the case for most transition metals 
and transition-metal oxides. In a density functional calculation with semi-local functionals, the spurious \ac{si} term acts effectively as a penalty term 
preventing electronic localization,  thus often spoiling 
agreement with experimental data for band gaps, 
magnetization, and other physical observables such as lattice constant and bulk modulus.

\subsection{LDA+U}
A popular way to reduce the \ac{si} 
while 
still remaining in the context of local or semi-local density functionals 
is by including in the kinetic functional \eqn{Eq_HK_nonint} also part of the electron-electron interaction, 
specifically the projection $\hHat$ of $\hVee$ on atomic-like orbital (se below).
The common choice is 
to consider only orbitals that are partially occupied within standard LDA, hence which suffer more from the \ac{si}. 
The non-interacting kinetic functional $\Ts[n(\br)]$ is thus turned into a modified kinetic functional $T_i[n(\br)]$:
\begin{align}\label{Eq_HK_modified}
\Ts[n(\br)] \ra T_i[n(\br)] = \min_{\Psio \ra n(\br)} \expect{\Psio}{\hT+\hHat}\,,
\end{align}
and the Hohenberg and Kohn functional changes into 
\begin{align}
\HKfunc[n(\br)] &= T_i[n(\br)] +\EH[n(\br)] \nonumber\\
& + \Exc[n(\br)] - \Edc[n(\br)]\,,\label{Eq_HK_plusU}
\end{align}
where $\Edc[n(\br)]$ is a double-counting energy which must cancel %to subtract  
the contribution of $\hHat$ already included within LDA.

In \eqn{Eq_HK_modified} the constrained-search is still restricted to the space of Slater-determinants 
$\Psio$, so that the modified kinetic functional can be dealt with within an independent-particle picture, 
and therefore included in the Kohn-Sham scheme. Essentially, the interaction $\hHat$ is treated by Hartree-Fock, 
which is devoid of SI -- while still unable to capture the Mott localization phenomenon, 
a correlation effect. 
In section \ref{Sect:gutz} we shall discuss how to improve the functional $T_i$ so as to make Mott physics 
accessible. Here in addition we  
briefly discuss how to define properly $\hHat$.
Typically $\hHat=\sum_{\bR} \hHat^{(\bR)}$, with $\hHat^{(\bR)}$ accounting for the leading order 
multipolar expansion of the Coulomb interaction projected onto a selected set of 
atomic-like orbitals $\ket{\phi^{(l)}_{\bR,m}}$ with angular momentum $l$  
at atomic site $\bR$ in the lattice, 
\begin{eqnarray}
\hat{H}^{(\bR)}_{\text{at}} &=& \frac{F_0}{2}\,\hat{N}_\bR\left(\hat{N}_\bR-1\right)
+ \frac{1}{2}\,\sum_{L>0}^{2l}F_L\,\left(C^{l0}_{l0\,L0}\right)^2\nonumber\\
&& ~~~~~~\sum_{M=-L}^L\,(-1)^M C^{lm}_{lm'\,LM}
C^{lm_1}_{lm'_1\,L-M}\nonumber\\
&& ~~~~~~~~~~~~~~c^\dagger_{\bR,m\sigma}c^\dagger_{\bR,m_1\sigma_1}
c^\dagga_{\bR,m'_1\sigma_1}c^\dagga_{\bR,m'\sigma}
\label{Eq:hHat_exact}
\end{eqnarray}
where $\hat{N}_\bR$ is the total electron number operator at site $\bR$ projected onto 
the selected set of atomic orbitals, $L = 2n$ with $n=1,\dots,l$, and $C^{lm}_{lm'\,LM}$ are 
the Clebsch-Gordan coefficients. The parameters $F_L$ are commonly 
known as Slater integrals. The first term on the right-hand side of \eqn{Eq:hHat_exact}, 
which we shall denote hereafter as $\hHub^{(\bR)}$, is a pure charge repulsion usually referred 
to as the Hubbard term, its coupling constant $F_0$ generally called the "Hubbard $U$". The remaining terms instead 
enforce Hund's first and second rules, hence they may be referred to as 
the Hund's rule exchange ($\hHund$).
In fact, in the case  of $p$ orbitals ($l=1$), the exact multipolar expansion can be rewritten solely in terms 
of the number operator $\hN_\bR$, the total spin $\mathbf{S}_\bR$ and total angular momentum  $
\mathbf{L}_\bR$ operators projected on the set $\ket{\phi^{(1)}_{\bR,m}}$:
\begin{align}
\label{Eq_Coulomb_on_p}
\hHat &= \frac{F_0}{2} \left[\hat{N}_\bR\left(\hat{N}_\bR-\mathds{1}\right)\right] \\
&+ \frac{F_2}{2}\left[\frac{4}{5}\hat{N}_\bR-\frac{\hat{N}_\bR^2}{5}-
\frac{3}{25}\left(4 \hat{\mathbf{S}}_\bR\cdot\hat{\mathbf{S}}_\bR  +\hat{\mathbf{L}}_\bR\cdot\hat{\mathbf{L}}_\bR\right)\right]\,,\nonumber
\end{align}
explicitly 
showing the content of the first two Hund rules. For $l>1$, it is no longer
possible to  
rewrite \eqn{Eq:hHat_exact} in terms of simple operators like spin and angular momentum.

The well-known \ac{ldau} method truncates the multipolar expansion of the Coulomb operator, 
\eqn{Eq:hHat_exact}, to the zeroth-order term, therefore setting $\hHat=\hHub$.
With this recipe, the density dependence of the expectation value $\expect{\Psio}{\hHat}$ can be written 
in terms of the matrix elements 
$\no_{lm\sigma\bR,lm'\sigma'\bR} = \expect{\Psio}{c^\dagger_{\bR, lm\sigma}
c^\dagga_{\bR,lm'\sigma'}}$ 
of the local single-particle density matrix $\hat{n}^{(0)}_\bR$ , which is an implicit function of the density $n(\br)$. 
If  lattice periodicity is unbroken and the set of correlated orbitals is characterized by a single 
value of the angular momentum, we can drop both indices $l$ and $\bR$ in any local operator, 
and write $\hat{n}^{(0)}_\bR = \hat{n}^{(0)}$,  $\forall \bR$.
The double-counting correction $\Edc[n(\br)]$ in \ac{ldau} is commonly chosen 
so as to cancel   $\expect{\Psio}{\hHat}$ in the limiting case 
of an idempotent single-particle density matrix $\hat{n}^{(0)}$,\cite{mazin} which corresponds to assuming 
that, within straight \ac{lda}, $\langle  \hat{n}^{(0)}_\bR \hat{n}^{(0)}_\bR \rangle = 
\langle  \hat{n}^{(0)}_\bR  \rangle\langle  \hat{n}^{(0)}_\bR \rangle $.
With this assumption, the $U$-dependent part of Kohn-Sham Hamiltonian is equal to 
the positive definite contribution
\begin{align}
\label{Eq:LDAU_total_ham}
\expect{\Psio}{\hHat}-\Edc[n(\br)] = \frac{U}{2} \text{Tr}\Big[\hat{n}^{(0)}\big(
1-\hat{n}^{(0)}\big)\Big].
\end{align}
An optimal value of $U$ can be estimated by linear response calculations \cite{Cococcioni_DeGironcoli_LinLDAU, 
Anisimov_Zaanen}, or 
empirically determined by agreement with  experimental data. 

The advantage of using \eqn{Eq:LDAU_total_ham} to improve the description of systems with 
strongly localized electrons is both its simplicity, involving no further computational effort than 
that needed to solve the Kohn-Sham equations, and its success in removing the self-interaction 
whenever $U$ is sensibly chosen.
However, there are of course situations in which the empirical \ac{ldau} functional will 
not be adequate.
We previously mentioned that Mott localization 
because of its genuinely many-body, collective nature,  
is not accessible by \ac{ldau} 
nor by any other technique that relies on a single-particle description.
Moreover, it is well known that only the spherically-averaged strength of the exchange-correlation 
hole is correctly accounted for by the LDA functional, but not its angular dependence. For these reasons 
one cannot expect that
\ac{ldau} will be apt to describe systems that display strongly orbital-dependent correlations, as was shown 
to be the case of body-centered cubic iron.\cite{anisimov} Indeed recent studies
on iron pnictides and chalcogenides\cite{Kotliar_coherence_incoherence, Kotliar_ironcalco, magnetism_chargedyn_pnictides}  
suggest that the orbital selectivity displayed by these iron compounds crucially depends on atomic Hund's rules. These 
observations indicate that a way to further improve LDA beyond \ac{ldau} will not only be the inclusion 
of correlations in the modified kinetic functional so as to make Mott localization accessible, 
but also 
the introduction of an appropriate expression for Hund's interaction $\hat{H}_{\text{Hund}}$ in 
the atomic Hamiltonian $\hat{H}_{\text{at}}$, so as to account for orbital selectivity. However, 
when Hund's rule exchange, the second term in 
the r.h.s. of \eqn{Eq:hHat_exact}, is taken into account, one faces the problem of finding a proper expression 
for electron double counting. The latter should by definition be equivalent to the LDA approximation 
to the atomic interaction energy, \eqn{Eq:hHat_exact}. However, that average depends in principle on the 
specific point symmetry of the system, and one cannot find a general expression valid for every case. 
The conventional way to proceed is to dismiss the hope of including within \ac{ldau} the whole 
atomic interaction \eqn{Eq:hHat_exact}, and instead be content with only terms that depend 
on angular-averaged local operators, specifically the total number operator $\hat{N}_\bR$ and total spin 
$\hat{\mathbf{S}}_\bR$. These terms are identified by noting that, using the 
re-coupling formula
\begin{eqnarray*}
&&\sum_M\,  (-1)^M\,  C^{lm}_{lm'\, LM} C^{lm_1}_{lm'_1\,L-M} = 
\sum_{\Lambda\lambda}\, (2\Lambda+1)\, 
\begin{Bmatrix}
L & l & l\\
\Lambda & l & l
\end{Bmatrix} \\
&&  ~~~~~~~~~~~~~~~~(-1)^{L+\Lambda}\, (-1)^{\lambda}\,C^{lm}_{lm'_1\,\Lambda\lambda} 
\, C^{lm_1}_{lm'\,\Lambda-\lambda}\, ,
\end{eqnarray*}
the $L>0$ contribution of \eqn{Eq:hHat_exact} can be also written as 
\begin{eqnarray*}
\hat{H}^{(\bR)}_{\text{at}\, L>0} &=& 
- \frac{1}{2}\sum_{mm_1m'm'_1}\sum_{\sigma\sigma_1}\sum_{L>0}^{2l}F_L\,\left(C^{l0}_{l0\,L0}\right)^2\nonumber\\
&& ~~\sum_{\Lambda}\sum_{\lambda=-\Lambda}^\Lambda 
(-1)^{\Lambda+\lambda} (2\Lambda+1) \begin{Bmatrix}
L & l & l\\
\Lambda & l &l 
\end{Bmatrix}\nonumber\\
&& C^{lm}_{lm_1'\,\Lambda\lambda}C^{lm_1}_{lm'\,\Lambda-\lambda}\,
c^\dagger_{lm\sigma}c^\dagga_{lm'_1\sigma_1}c^\dagger_{lm_1\sigma_1}c^\dagga_{lm'\sigma}\,,
\end{eqnarray*}
where $\{\dots\}$ denote the Wigner $6j$-symbols. We can then select out the term with $\Lambda=0$, 
which depends on rotationally invariant densities, 
re-couple back $m$ with $m'$ and $m_1$ with $m'_1$ in the remaining terms, and iterate the procedure.
At the end, we obtain a term that involves rotationally invariant densities, plus another interaction that 
cannot be expressed by any means in terms of those densities. The former 
together with the Hubbard $U$ define the part of the atomic interaction \eqn{Eq:hHat_exact} 
easier to implement within \ac{ldau}, namely  
\begin{align}
\hat{H}^{(\bR)}_{\text{at}} &\simeq& \frac{U}{2} \hat{N}_\bR\Big(\hat{N}_\bR-1\Big)
- \frac{2l+1}{2l+2}\,J \,\Bigg[\hat{\mathbf{S}}_{\bR}\cdot \hat{\mathbf{S}}_{\bR}
-\frac{3}{4}\,\hat{N}_\bR  \nonumber\\
&& + \fract{\hat{N}_\bR\left(\hat{N}_\bR-1\right)}{4} 
+\fract{\hat{N}_\bR\left(\hat{N}_\bR-1\right)}{2(2l+1)} 
\Bigg],\label{Eq:hHat_approssimata}
\end{align}
where $J$ is conventionally defined as \cite{Cococcioni_DeGironcoli_LinLDAU, Anisimov_Zaanen} 
\be
J = \frac{1}{2l}\sum_{L>0}^{2l} \, \left(C_{l0\, L0}^{l0}\right)^2\, F_L,\label{def:J}
\ee
which, for $d$-orbitals, i.e. $l=2$, is $J=(F_0+F_4)/14$. The double counting term associated with 
\eqn{Eq:hHat_approssimata} is obtained  analogously as before and reads in the general case of a 
spin-polarized calculation 
\begin{align}
E_{\text{dc}} &=& \frac{U}{2} N\big(N-1\big) - \frac{2l+1}{4l+4}\,J \,\bigg[ N_\up\left(N_\up-1\right) 
\nonumber\\
&& ~~~~~~~~~~+ 
 N_\dw\left(N_\dw-1\right) + \fract{N(N-1)}{2l+2}\bigg].\label{double-counting-Hund}
\end{align}
The expression \eqref{Eq:hHat_approssimata} can be further simplified to get rid of the $l$ dependence, 
by readsorbing the $l$ in the definition of $J$, and by adopting a simplified version
of the last term in square brackets, leading to the following results
\begin{align}\label{Eq:recast_J}
\hHund=-J \left\{ \hat{S}^2 - \frac{3}{4} \hat{N} + \frac{\hat{N} (\hat{N}-1)}{4}  +\sum_{m} \hat{n}_{m\up}\hat{n}_{m{\dw}} \right\}\,.
\end{align}
for which we choose a double-counting 
energy 
of the type
\begin{align}\label{Eq:Edc_ours}
&\Edc^{\rm Hund}[n(\br)] = \nonumber \\
&= -J \sqbra{\frac{\Nup (\Nup-1)}{2} +\frac{\Ndw(\Ndw-1)}{2} + \frac{\Nup\Ndw}{2l+1} }\,.
\end{align}
\subsection{Extending LDA+U to LDA+Gutzwiller}\label{Sect:gutz}
The key difference between \ac{ldag} and \ac{ldau} resides in the definition of the modified kinetic 
functional $T_{\rm i}$. Within \ac{ldag}, the definition \eqn{Eq_HK_modified} changes to
\begin{align}\label{Eq_GW_kinetic}
T_{\rm i}[n(\br)] \ra T_{\rm G}[n(\br)] = \min_{\PsiG \ra n(\br)} \expect{\PsiG}{\hT+\hHat}\,,
\end{align}
where the  wavefunction $\kPsiG$ is defined as
\begin{align}\label{Eq_Gutzwav_def}
\PsiG = \Gpg{} \kPsio = \prod_{\bR} \Gpl{\bR} \kPsio\,.
\end{align}
In the above equation, $\kPsio$ is still a Slater determinant, and the elements of novelty are 
the operators $\Gpl{\bR}$, which are linear transformations acting on the configurational space of 
a chosen set of local orbitals at lattice site $\bR$. As in \ac{ldau}, this set of orbitals $\phi_{m,\bR}$ 
retain well defined atomic angular momentum $l$, $m$ being its projection on a given quantization axis.  
The operator $\Gpl{\bR}$ can be generally written as
\begin{align}\label{Eq_multiband_projector}
\Gpl{\bR} = \sum_{\Gamma \Gamma'} \Lam_{\Gamma\Gamma',\bR}\,
\ket{\Gamma,\bR}\bra{\Gamma',\bR}\,,
\end{align}
where $\ket{\Gamma,\bR}$ denote many-body configurations of electrons occupying the 
orbitals $\phi_{m,\bR}$. Differently from \ac{ldau}, the expectation value of the kinetic plus atomic 
interaction operators will not depend solely on the Slater determinant $\kPsio$,  but also on the variational 
parameters $\Lam_{\Gamma\Gamma',\bR}$ that define $\Gpl{\bR}$. 

Computing exact expectation values on the Gutzwiller wavefunction for lattices of finite coordination 
is a task that can be accomplished only numerically, e.g. through Variational Quantum Monte Carlo.\cite{Sorella_SR,Sorella-VMC} 
For infinite-coordination lattices, an exact expression can be instead computed analytically. There is in fact 
a close connection between the Gutzwiller variational approach in the limit of infinite lattice coordination  
and dynamical mean field theory.\cite{DMFT} In that limit, the single particle self-energy matrix 
$\Sigma(\epsilon,\bk) = \Sigma(\epsilon)$ becomes purely local, hence momentum independent. 
DMFT allows to evaluate exactly $\Sigma(\epsilon)$ by solving an auxiliary Anderson impurity model 
constructed in such a way as to have the same self-energy. The Gutzwiller variational approach is instead 
a consistent approximation to the exact solution, which assumes a Fermi-liquid expression 
$\Sigma(\epsilon) \simeq \Sigma(0) + \left(1-Z^{-1}\right)\epsilon$, where $Z$ is commonly refereed 
to as the quasiparticle weight. Because of this assumption, the Gutzwiller wavefunction can describe only states 
whose elementary excitations are quasiparticles, %hence 
such as 
Landau-Fermi liquids and insulators that can be 
represented through a Slater determinant. However, the additional freedom brought by the parameter $Z$, 
whose value is strictly 
$Z = 1$ within Hartree-Fock 
and in \ac{ldau}, opens the possibility to access  
strongly correlated metals, $Z\ll 1$,  
and thus the approach to a  
Mott transition, where $Z\to 0$. 
Although DMFT is exact only in the limit of infinite coordination, it is currently used as an approximation 
in realistic finite-coordination lattices, under the hypothesis that (strong) correlation effects beyond Hartree-Fock (HF) are well represented by 
$\Sigma(\epsilon,\bk) \simeq \Sigma_{\text{HF}}(\bk) + \Sigma(\epsilon)$, where 
$\Sigma_{\text{HF}}(\bk) $ is the HF self-energy, eventually including frequency-dependent 
random-phase-like contributions,\cite{LDA+cRPA+DMFT} and the correction $\Sigma(\epsilon)$ is momentum independent and can be obtained by DMFT. Under the same assumptions, one can keep using the 
formal results of the Gutzwiller variational approach, that are strictly valid only in infinite-coordination lattices, also in finite-coordination ones, an approximation refereed to as the {\sl Gutzwiller approximation} (\acs{ga}). 
In other words, the GA should be better regarded as an approximation to DMFT, when either of them are used in finite-coordination lattices, rather than an approximation to the exact evaluation of average values on the Gutzwiller wavefunction, 
\eqn{Eq_Gutzwav_def}. This viewpoint, which we underwrite, is our motivation for adopting the Gutzwiller 
approximation in combination with \ac{ldau} as an alternative to \ac{ldadmft}, 
at the cost of less rigor, but as we shall show with gain in simplicity and flexibility. 
\subsubsection{Expectation values in the Gutzwiller Approximation}\label{Sect:Gutz_expect}
In order to  determine 
the functional $T_{\rm G}[n(\br)]$, one should be able to compute expectation values 
of both many-body on-site operators such as those contained in $\hHat$, and off-site single-particle 
operators, which are present in the definition of the kinetic operator $\hat{T}$. In all what follows, 
we shall use the formalism presented in Ref.~\onlinecite{BaroneLanata}. 

First of all, the Slater determinant $\mid\Psi_0\rangle$ defines the uncorrelated one-body local 
density-matrix $\hat{n}^{(0)}_\bR$ (the same matrix that enters the \ac{ldau} energy 
correction term \eqn{Eq:LDAU_total_ham}), with elements 
\be
\no_{\bR m\sigma,\bR m'\sigma'} = \langle\Psi_0\mid 
c^\dagger_{\bR,m\sigma} c^\dagga_{\bR,m'\sigma'}\mid\Psi_0\rangle, \label{II.B-n0}
\ee
where 
$c^\dagger_{\bR,m\sigma}$ 
creates a spin-$\sigma$ electron in orbital $\phi_{m,\bR}$. 
$\hat{n}^{(0)}_\bR$ is diagonalized by a unitary transformation that turns the 
original basis of operators 
$c^\dagger_{\bR, m\sigma}$ 
into the natural basis of operators 
$c^\dagger_{\bR, \gamma\sigma}$
, assuming 
invariance with respect to spin rotations around the $z$-axis. In the natural basis, the one-body 
density matrix is therefore diagonal, with eigenvalues $\nnat_{\bR,\gamma\sigma}$. 
In the natural-orbital Fock basis, with states 
\[
\mid \{n_{\bR,\gamma\sigma}\}\rangle 
\equiv \prod_{\gamma\sigma}\,\left(c^\dagger_{\bR,\gamma\sigma}\right)^{n_{\bR,\gamma\sigma}}\,
\mid 0\rangle,
\] 
it follows that the probability matrix 
\begin{eqnarray}
&&P^{(\bR)}_{0,\{n_{\bR,\gamma\sigma}\}\{m_{\bR,\gamma\sigma}\}} \equiv 
\langle \Psi_0\mid \, 
\mid \{m_{\bR,\gamma\sigma}\}\rangle\langle \{n_{\bR,\gamma\sigma}\}\mid\, 
\mid \Psi_0\rangle \nonumber\\
&&~~~~~~~~~=
P^{(\bR)}_{0,\{n_{\bR,\gamma\sigma}\}}\,\delta_{\{n_{\bR,\gamma\sigma}\}
\{m_{\bR,\gamma\sigma}\}} \nonumber\\
&& ~~~~~~~~=\prod_{\gamma\sigma} \left(\nnat_{\bR,\gamma\sigma}\right)^{n_{\bR,\gamma\sigma}}\,
\left(1-\nnat_{\bR,\gamma\sigma}\right)^{1-n_{\bR,\gamma\sigma}}\,, \label{Eq_pizero_explicit}
\end{eqnarray}
is diagonal, too. It is actually convenient\cite{BaroneLanata} to rewrite the operator \eqn{Eq_multiband_projector} in a mixed basis representation
as 
\begin{align}\label{IIB.Eq_multiband_projector}
\Gpl{\bR} = \sum_{\Gamma \{n_{\bR,\gamma\sigma}\}} \left(
\fract{\Phi_{\Gamma \{n_{\bR,\gamma\sigma}\},\bR} }{P^{(\bR)}_{0,\{n_{\bR,\gamma\sigma}\}}}\right)
\ket{\Gamma,\bR}\bra{\{n_{\bR,\gamma\sigma}\}}\,,
\end{align}
where $\mid \Gamma,\bR\rangle$ is a state, e.g. a Fock state, 
in the original basis, whereas $\mid \{n_{\bR,\gamma\sigma}\}\rangle$ is a Fock state in the natural basis. This mixed representation 
simplifies considerably the calculations. 
In order to use the Gutzwiller approximation, we  
need to impose the two following 
constraints on the matrix $\hat{\Phi}_\bR$ with elements $\Phi_{\Gamma \{n_{\bR,\gamma\sigma}\},\bR}$:\cite{BaroneLanata}
\begin{align}
\Tr{\hPhigd_{\bR}\hPhig_{\bR}} &= 1\,, \label{Eq_Gw2_constr1}\\
\Tr{\hPhigd_{\bR}\hPhig_{\bR}\ocd_{\bR, \gamma\sigma}\oc_{\bR, \gamma'\sigma'}} &= 
\nnat_{\bR,\gamma\sigma}\delta_{\gamma\gamma'}\,\delta_{\sigma\sigma'}\,,\label{Eq_Gw2_constr2}
\end{align}
where $\ocd_{\bR, \gamma\sigma}$ is the matrix representation of the Fermi operator in its Fock basis. 
If these constraints are fulfilled, then within the Gutzwiller approximation, which we recall is exact for infinite-coordination lattices,
we have
\begin{align}\label{Eq:renorm_onsite}
\expect{\PsiG}{\hat{O}_{\bR}}= \Tr{\hPhigd_{\bR}\hat{O}_{\bR}\hPhig_{\bR}}\,,
\end{align}
where $\hat{O}_{\bR}$ is the matrix representation of any local operator. 
The inter-site density matrix can be computed from
\begin{align}\label{Eq:renorm_offsite}
\expect{\PsiG}{\ocd_{\bR,m\sigma}\oc_{\bR,m'\sigma}} = 
&\sum_{\gamma\gamma'}\, R_{\gamma m;\sigma,\bR}^\dagger \,R_{m'\gamma';\sigma,\bRp}^\dagga\nonumber\\
& \expect{\Psio}{c^\dagger_{\bR,\gamma\sigma'}c^\dagga_{\bRp,\gamma'\sigma'}} \,,
\end{align}
where
\begin{align}\label{Eq_Rparam2} 
R^{\dagger}_{\gamma m,\sigma,\bR} = \fract{\Tr{\hat{\Phi}_\bR^\dagger\,\ocd_{\bR, m \sigma}\,
\hat{\Phi}_\bR^\dagga\,\oc_{\bR, \gamma \sigma}}}{\sqrt{\nnat_{\bR,\gamma\sigma} (1-\nnat_{\bR,\gamma\sigma})}} \,,
\end{align}
can be regarded as a wavefunction renormalization matrix. 
Here $\ocd_{\bR, m \sigma}$ is the matrix representation of the original operators in the basis of states 
$\mid \Gamma,\bR\rangle$. When this is the Fock basis constructed by the same original operators, 
their matrix representation is actually independent of the basis of single-particle wavefunctions which 
they refer to, hence it is the same as for the  $\ocd_{\bR, \gamma \sigma}$ operators of the natural basis. 
In reality, in most cases that are relevant for real materials the natural basis that diagonalizes the local density 
matrix is determined fully by the lattice symmetry, hence it is possible and convenient to write the Hamiltonian 
directly in that basis. In the above formulas, this corresponds to identifying the set of labels 
$\{m\}$ with $\{\gamma\}$.  Since the natural basis is such both for the uncorrelated on-site density matrix 
\begin{align}
\no_{\bR m\sigma,m'\sigma'} &= \langle\Psi_0\mid 
\ocd_{\bR,m\sigma} \oc_{\bR,m'\sigma'}\mid\Psi_0\rangle \nonumber\\
&=  \Tr{\hat{\Phi}_\bR^\dagger\, \hat{\Phi}_\bR^\dagga\,\ocd_{\bR, m \sigma}\oc_{\bR, m' \sigma}
} \nonumber\\
&= \delta_{mm'}\, \no_{\bR, m\sigma},\label{II.B-density-matrix-var}
\end{align}
and for the correlated one
\begin{align}
n_{\bR m\sigma,m'\sigma'} &= \langle\Psi_{\text{G}}\mid 
\ocd_{\bR,m\sigma} \oc_{\bR,m'\sigma'}\mid\PsiG\rangle \nonumber \\
&= \Tr{\hat{\Phi}_\bR^\dagger\, \ocd_{\bR, m \sigma}\oc_{\bR, m' \sigma}\, 
\hat{\Phi}_\bR^\dagga} \nonumber\\
&=
\delta_{mm'}\, n_{\bR, m\sigma},\label{II.B-density-matrix-true}
\end{align}
generally with different eigenvalues, it is not difficult to realize that the wavefunction renormalization 
matrix \eqn{Eq_Rparam2} becomes diagonal, i.e. 
\begin{align}\label{Eq_Rparam2-1} 
R^{\dagger}_{m' m,\sigma,\bR} = \fract{\Tr{\hat{\Phi}_\bR^\dagger\,\ocd_{\bR, m \sigma}\,
\hat{\Phi}_\bR^\dagga\,\oc_{\bR, m' \sigma}}}{\sqrt{\nnat_{\bR,m'\sigma} (1-\nnat_{\bR,m'\sigma})}} 
= \delta_{mm'}\,R_{m\sigma,\bR}^\dagger\,.
\end{align} 

The Eqs.~(\ref{Eq:renorm_onsite})--(\ref{Eq_Rparam2-1}) are the basic formulas that allow to 
evaluate the average value of the Hamiltonian as a functional of the Slater determinant and of the matrices 
$\hat{\Phi}_\bR$, hence to solve the variational problem.
\section{The Gutzwiller functional in practice}\label{Sec:Gutz_practice}
In this section we show how to perform a density-self-consistent \ac{ldag} calculation on a realistic system, 
namely %the body-centered cubic iron that,
$bcc$ Fe which,
as mentioned in the Introduction, although a %relatively 
basic and supposedly simple system, still exhibits controversial aspects. 

We first have to select the {\sl correlated} orbitals to be treated by the Gutzwiller operator. In the present 
case the choice is simple: the $3d$ orbitals of Fe. %In addition, this example 
This case 
is one of those mentioned earlier 
in which the natural basis is determined by symmetry and corresponds to the cubic crystal field split 
$d$ orbitals, namely the $e_g$ doublet and the $t_{2g}$ triplet. In this representation the 
formulas Eqs.~(\ref{II.B-density-matrix-var})--(\ref{Eq_Rparam2-1}) hold, which is a great simplification. 
Furthermore, since $bcc$ is a Bravais lattice, the positions $\bR$ of Fe atoms also label unit cells, hence 
by translational symmetry we can safely assume that the variational matrix parameters $\hat{\Phi}_\bR 
= \hat{\Phi}$ are independent of $\bR$. So are therefore the eigenvalues of the local density matrices, 
$\no_{\bR,m\sigma} = \no_{m\sigma}$ and $n_{\bR,m\sigma}=n_{m\sigma}$, as well as the wavefunction 
renormalization $R_{m\sigma,\bR}=R_{m\sigma}$. To lighten notations, in what follows the orbital labels 
$m$ will refer both to the correlated set and to the uncorrelated ones, unaffected by the action of the Gutzwiller operator. In the last paragraph of this section we shall come back to this point.   

We define the Gutzwiller density functional as
\begin{align}\label{Eq:F_functional-1}
&{\cal F}[n(\br)] = \min_{\PsiG \ra n(\br)} {\cal E}[\PsiG,n(\br)]\,.
\end{align}
where the quantity ${\cal E}[\PsiG,n(\br)]$ undergoing constrained minimization is
\begin{align}\label{Eq:E_functional}
&{\cal E}[\PsiG,n(\br)] = \expect{\PsiG}{\hT+\hHint}+\nonumber\\&+\int \Vext(\br) n(\br) \de{\br}+\tilde{E}_{\rm H}[n(\br)] + \tilde{E}_{\rm xc}[n(\br)] - \Edc[n(\br)]\,.
\end{align}
For our purposes, it is convenient to rewrite \eqn{Eq:F_functional-1} as a minimization constrained 
with respect to the ``uncorrelated'' density $\no(\br)$,
\begin{align}\label{Eq:F_functional}
&{\cal F}[\no(\br)] = \min_{\Gpl{},\Psio \ra \no(\br)} {\cal E}[\Psio,\Gpl{},\no(\br)]\,,
\end{align}
where ${\cal E}[\Psio,\Gpl{},\no(\br)] = {\cal E}[\PsiG(\Psio,\Gpl{}),n(\Psio,\Gpl{})]$.
The dependence of the ``correlated'' density $n(\br)$ upon the ``uncorrelated'' density $\no(\br)$ can 
be made explicit once one writes them in terms of the one-body ``correlated'' density-matrix of the periodic system
\begin{align}\label{Eq:1b_corr_densmat}
\D_{mm',\sigma,\bR} = \langle \Psi_{\text{G}}\mid \ocd_{\bR,m\sigma}
\oc_{\bm 0,m'\sigma}\mid\Psi_{\text{G}}\rangle,
\end{align}
and of the ``uncorrelated'' density-matrix 
\begin{align}\label{Eq:1b_uncorr_densmat}
\Do_{mm',\sigma,\bR}= \langle \Psi_{0}\mid \ocd_{\bR,m\sigma}
\oc_{\bm 0,m'\sigma}\mid\Psi_{0}\rangle,
\end{align}
namely
\begin{align}
\no(\br) &= \sum_\sigma \, \no_\sigma(\br) \nonumber\\
&= \sum_{m,m',\sigma, \bR} \Do_{mm',\sigma,\bR}\, 
\phi^\ast_{m,\bR}(\br) \,
\phi_{m',\bm 0}(\bm{r})\, , \label{Eq_dens_uncorr}\\
n(\bm{r}) &= \sum_\sigma \, n_\sigma(\br) \nonumber\\
&= 
\sum_{m,m',\sigma, \bm R} \D_{mm',\sigma,\bR} \, \phi^\ast_{m,\bR}(\br) \,
\phi_{m',\bm 0}(\bm{r})\, .
\label{Eq_dens_corr}
\end{align}
Indeed, $\D_{mm',\sigma,\bR}$ can be obtained by  $\Do_{mm',\sigma,\bR}$ using the recipe of the Gutzwiller Approximation:
\begin{align}\label{Eq_rendensmat}
\D_{mm',\sigma,\bR} = \begin{cases}
R_{m\sigma}^\dagger\, 
\Do_{mm',\sigma,\bR} \,R_{m'\sigma}\,, & \hspace{-0.3cm}\bR\neq \bm 0\,,\vspace{0.3cm}\\
\Tr{\hPhigd \,\hat{n}_{mm',\sigma}\,\hPhig\hspace{-0.2cm}}= 
\delta_{mm'} n_{m\sigma}\,,
& \hspace{-0.3cm}\bR = {\bm 0}\,,
\end{cases}
\end{align}
where $\hat{n}_{mm',\sigma}$ is the matrix representation on the local Fock space at site $\bR$ of 
$\ocd_{\bR,m\sigma}\ocd_{\bR,m'\sigma}$, which is independent of $\bR$ for a periodic system, 
and where $n_{m\sigma}$ is equal to $n_{\bR=0,m\sigma}$ defined in \eqn{II.B-density-matrix-true}.

In order to write ${\cal E}[\Psio,\Gpl{},\no(\br)]$ explicitly in terms of the new variables, we start from 
the first and second terms of \eqn{Eq:E_functional}.
We can now treat the kinetic and the external potential terms on the same footing through 
\begin{align}\label{Eq_kinextgutz_trace}
& \expect{\PsiG}{\hT} + \int n(\br) \Vext(\br) d\br = \nonumber\\
& \sum_{m,m',\sigma,\bR} \Big(T_{mm',\bR} + \Vextm_{mm',\bR}\Big)
\,\D_{mm',\sigma,\bR}\,,
\end{align}
where values of $T_{mm',\bR}$ and $\Vextm_{mm',\bR}$ are the  spin-independent matrix elements of 
the kinetic and external potential operators computed between our basis orbitals at sites $\bR$ and $\bm 0$, 
i.e. 
\begin{align}
\Vextm_{mm',\bR} = \int \phi^\ast_{m,\bR}(\br) \Vext(\br) 
\phi_{m',\bm 0}(\br)d \br\,,\\
T_{mm',\bR} = -\frac{\hbar^2}{2m} \int \phi^\ast_{m,\bR}(\br) 
\Big[\nabla^2 \phi_{m',\bm 0}(\br)\Big] d \br\,
\end{align}
and compute the value of the atomic interaction energy $\expect{\PsiG}{\hHat}$ using the Gutzwiller Approximation recipe
\begin{align}\label{Eq:eat}
\Eat[\Psio,\Gpl{}] = \expect{\PsiG}{\hHat} = \Tr{\hPhigd \hHat \hPhig}
\end{align}
In order to simplify the density self-consistent \ac{ldag} minimization we decided to use the
Hartree $\tilde{E}_{\text{H}}[n(\br)]$ and exchange-correlation $\tilde{E}_{\text{xc}}[n(\br)]$ 
functionals as the \ac{lda} functionals linearized around the uncorrelated density $\no(\br)$. 
We checked {\sl a posteriori} the accuracy of such a linearization. 
The modified Hartree functional then reads
\begin{align}\label{Eq:Hartree1}
\tilde{E}_{\rm H}\left[\no(\br),n(\br)\right] &\simeq \frac{e^2}{2}\int d\br d\brp\; \fract{\no(\br) \no(\brp)}{|\br - \brp|}
\nonumber \\
&~~~+ \int d\br \,\delta n(\br)\,\vH[\no(\br)]\,,
\end{align}
where $\delta n(\br) = \sum_\sigma \delta n_\sigma(\br)=\sum_\sigma n_\sigma(\br)-\no_\sigma(\br)$ 
and $\vH[\no(\br)]$ is the conventional Hartree potential, whereas the exchange-correlation functional is 
\begin{align}\label{Eq:exc1}
\tilde{E}_{\rm xc}\left[\no(\br),n(\br)\right] &=  \sum_\sigma\, \int \de{\br} \no_\sigma(\br)\, 
\epsilon_{\text{xc},\sigma}[\no(\br)] 
\nonumber\\
&~~+  \int d\br \, v_{\text{xc},\sigma}[\no(\br)] \, \delta n_\sigma(\br)\,,
\end{align}
$\vxc[\no(\br)]$ being the \ac{lda} exchange-correlation potential.
Note that the choice of $\tilde{E}_{\rm H}$ involves neglecting a term
\begin{align}
\Delta\EH\left[\no(\br),n(\br)\right] &= \tilde{E}_{\rm H}\left[\no(\br),n(\br)\right]-\EH[n(\br)] =\nonumber\\
&=\frac{e^2}{2}\int d\br d\brp\, \fract{\delta n(\br) \delta n(\brp)}{|\br - \brp|}
\end{align}
which can be interpreted as the energy of correlation-induced charge fluctuations. This term, together with 
the corresponding one neglected for the exchange-correlation functional, $\Delta\Exc\left[\no(\br),n(\br)\right]$, 
can be computed at the end of the \ac{ldag} calculation in order to provide an estimate of the error due to 
approximations~\eqref{Eq:Hartree1} and~\eqref{Eq:exc1} (see \rtab{Tab:delta_energies}).
It is worth mentioning that the linearization~\eqref{Eq:exc1} of exchange-correlation energy around 
the ``uncorrelated'' density does not spoil the sum rule for the \ac{lda} exchange-correlation hole.
As for the double-counting term, similarly to what is done within \ac{ldau}, it is chosen as a function 
of the local ``uncorrelated'' density-matrix $\no$ only, $\Edc[n(\br)] = \Edc[\no]$. In \sect{Sec:results} we take
as its explicit form the one of \eqn{double-counting-Hund}, having chosen our atomic interaction Hamiltonian $\hHat$
to be the expression of \eqn{Eq:hHat_approssimata}.

\subsection{Three-step minimization of the LDA+Gutzwiller functional}\label{Sec:threestep_mini}

The two densities $n(\br)$ and $\no(\br)$ must be such that Gutzwiller constraints are fulfilled. 
In our case where original and natural basis coincide, the constraints on the density matrix can be written as
\begin{align}
\Do_{mm', \sigma,\bR= \bm 0} &= \nnat_{m\sigma}\,\delta_{mm'}\,,\label{Eq_df2_gwconst1}\\
\Tr{\hPhigd\hPhig \hat{n}_{mm',\sigma}} &= \nnat_{m\sigma}\,\delta_{mm'}\,,\label{Eq_df2_gwconst2}
\end{align}
where we regard $\nnat_{m\sigma}$ as an additional independent variational parameter of the density 
functional. These constraints can be enforced with Lagrange multipliers, together with the first Gutzwiller constraint 
\begin{align}
\Tr{\hPhigd\hPhig} &= 1\,.\label{Eq_df_gwconst}
\end{align}

Summing up all contributions and adding the electrostatic ion-ion interaction $E_{\rm ion}$, 
we find that the overall functional we need to minimize has the form
\begin{widetext} 
\begin{align}\label{Eq_Gwdensfunc_full}
\DF\Big[n(\br),\no(\br),\nnat_{m\sigma}\Big] &= \max_{\lambda\lambda'\lambda_0} 
\Bigg[{\cal K}[n(\br)] +\Eat[n(\br)] -\Edc[\nnat_{m\sigma}]+\EHo[\no(\br)] + \Exco[\no(\br)]
- \lambda_0 \left(\Tr{\hPhigd\hPhig} - 1\right)\nonumber \\
& -\sum_{mm'\sigma}\, 
\lambda'_{mm',\sigma}\left(\Do_{mm',\sigma, \bR=\bm 0}-\nnat_{m\sigma}\delta_{mm'}\right) 
- \lambda_{mm',\sigma} \left(\Tr{\hPhigd\hPhig \hat{n}_{mm',\sigma}}
-\nnat_{m\sigma}\delta_{mm'}\right)
\Bigg] +E_{\rm ion},
\end{align}
\end{widetext}
where the functional ${\cal K}[n(\br)]$ contains all terms which depend on $n(\br)$ linearly through 
the renormalized density matrix $\D$, namely
\begin{align}
{\cal K}(\D) &= \sum_{mm',\sigma,\bR} \bigg[T_{mm',\bR}+\VHm_{mm',\bR}
+
\Vxcm_{mm',\sigma,\bR}\nonumber\\
&~~~~~~~~~~~~~~~~~~~~~~~~~ +\Vextm_{mm',\bR}\bigg]\, \D_{mm',\sigma,\bR}\nonumber\\
&\equiv \sum_{mm',\sigma,\bR} \, {\cal K}_{mm',\sigma,\bR}\, \D_{mm',\sigma,\bR}\,,
\end{align}
where $\VHm_{mm',\bR}$ and $\Vxcm_{mm',\sigma,\bR}$ are the matrix elements of $\vH$ and $\vxc$ 
between the basis orbitals. 
For every fixed value of $\nnat_{m\sigma}$, we can optimize $\DF$ with respect to the two 
densities $\no(\br)$ and $n(\br)$. In practice, by inspection of equations~\eqref{Eq_dens_uncorr}, 
\eqref{Eq_dens_corr} and \eqref{Eq_rendensmat} one can see that this is equivalent to a minimization 
with respect to the Slater determinant $\ket{\Psio}$ and the \xGps\ contained in the operator $\hPhig$.
This minimization can be carried out in two separate steps:
\begin{enumerate}
\item first carry out a Siesta self-consistent calculation to find the Slater determinant $\Psio$ 
that optimizes $\DF[n(\br),\no(\br), \nnat_{m\sigma}]$ with respect to $\no(\br)$, enforcing the 
constraint~\eqref{Eq_df2_gwconst1} through an Augmented Lagrangian Method~\cite{Fletcher}.
The \xGps, and therefore the hopping renormalization parameters $R_{m\sigma}$, are kept fixed throughout 
this optimization. The atomic energy $\Eat[n(\br)]$ does not change, nor does the double-counting energy 
$\Edc[\no(\br)]$, which is a function of $\no(\br)$ only through $\nnat_{m\sigma}$.
The self-consistent single-particle Kohn-Sham equations allowing the minimization with respect to $\kPsio$ are
\begin{align}
\sum_{m'\bR}\,\mathcal{H}_{mm',\sigma,\bR}\,\psi_{m'\sigma,\bR} &= \varepsilon\, \psi_{m,\sigma,\bm0}\,,
\end{align}
where
\[
\mathcal{H}_{mm',\sigma,\bR} = {\cal K}_{mm',\sigma,\bR}+\Vo_{mm',\sigma,\bR} - \lambda'_{mm',\sigma}
\delta_{\bR\bm 0},
\]
and 
\begin{align}
\Vo_{mm',\sigma,\bR} &= \int \de{\br} \phi^\ast_{m,\bR}(\br)\Big\{\vH[\no(\br)] \nonumber\\
&~~~~~~~~~+ \vxc[\no(\br)]\Big\}\phi_{m',\bm 0}(\br)\,.
\end{align}
\item next, optimize $\DF$ with respect to \xGps\ by a Lanczos-improved \ac{lm} algorithm (see \appendixx{lm}), 
enforcing the constraints~\eqref{Eq_df2_gwconst2} and \eqref{Eq_df_gwconst}. During this optimization, 
only the term ${\cal K}[n(\br)]$ and the atomic energy $\Eat[n(\br)]$ in \eqn{Eq_Gwdensfunc_full} are modified.
These two quantities, together with the terms enforcing constraints for \xGps, build a quartic functional 
$F_{\hPhig}$ of the matrices $\hPhig$, with explicit form 
\begin{widetext}
\begin{align}\label{Eq:Gutzfunc_LDA}
F_{\hPhig} = \sum_{m,m',\sigma} \Bigg[&{\cal K}_{mm',\sigma,\bR=\bm 0}\,
\Tr{\hPhigd\hn_{mm',\sigma}\hPhig} + R_{m\sigma}^\dagger\tau_{mm',\sigma}
R_{m'\sigma} + 
\Tr{\hPhigd\hHat\hPhig}  \nonumber \\
& - \lambda_{mm',\sigma} \left(\Tr{\hPhigd\hPhig \hn_{mm',\sigma}}
-\nnat_{m\sigma}\delta_{mm'}\right) - \lambda_0 \left(\Tr{\hPhigd\hPhig} - 1\right)\Bigg]\,,
\end{align}
\end{widetext}
where $\tau_{mm',\sigma}$ is 
\begin{align}
\tau_{mm',\sigma} =  \sum_{\bR\neq 0, mm'} {\cal K}_{mm',\sigma,\bR}\, 
\Do_{mm',\sigma,\bR}\,.\label{Eq:hopping_matrix}
\end{align}

\end{enumerate}
These two steps are repeated one after the other until self-consistency is achieved over both densities $n(\br)$ and $\no(\br)$. 
Once converged, we are left with a total energy functional depending on the diagonal matrix 
elements $\nnat_{m\sigma}$, and that can be optimized  by steepest descent, so as to fulfill the stationary equations
\begin{align}
\frac{\partial{\cal K}[n(\br)]}{\partial \nnat_{m\sigma}} - \frac{\partial \Edc[\no(\br)]}
{\partial \nnat_{m\sigma}} + \lambda_{mm,\sigma}+\lambda'_{mm,\sigma} = 0\,.
\end{align}
The terms appearing in the above equations are the only ones depending on the local uncorrelated density 
matrix $\nnat_{m\sigma}$. The double-counting energy is a function of this density matrix only, while 
the functional ${\cal K}$, containing the renormalized density matrix $\D_{mm',\sigma,\bR}$, depends 
on $\nnat_{m\sigma}$ through the wavefunction renormalization parameters $R_{m\sigma}$.

\subsection{Atomic basis set angular momentum dependent renormalization for transition metals}

Equations~\eqref{Eq:1b_corr_densmat} and~\eqref{Eq:1b_uncorr_densmat} describe a density matrix 
on a basis of orthogonalized atomic orbitals. 
For a system described by a single set of atomic $d$-orbitals, the indices $m$ and $m'$ are allowed 
to run on every value of the magnetic quantum number, $ m=\{-2,-1,0,1,2\} $.  For a cubic system, 
the basis which diagonalizes the one-body density matrix 
$\D_{\bR,mm'}$ for $\bR=0$ is the basis of $d$ orbitals which have real harmonics as their angular part, 
so that $m=\{r^2-3z^2,x^2-y^2,xy,xz,yz\}$, or in the language of group representations, 
$m=\{e^{(1)}_{g}, e^{(2)}_{g},t^{(1)}_{2g}, t^{(2)}_{2g},t^{(3)}_{2g}\}$.
In general, simulating the electronic structure of a transition metal with an atomic basis set,  such as 
the Siesta code does, will require also $s$ orbitals to be present in the set, as well as $p$ orbitals in 
order to allow for polarization of $d$ and $s$ atomic orbitals due to the cubic crystal field. We need 
therefore to reframe \eqn{Eq_rendensmat} by adding an additional couple of indices $l=0,1$ besides $l=2$. Since 
we assume for simplicity that Gutzwiller renormalization affects only $d$-type orbitals, we have that

%%%%%%%%%%%%%%%%%%%%%%%%%%%%%%%%%%%%%%%%%%%%%%%%%%%%%%%%%%%%%%%%%%%%%%%%

\begin{align}\label{Eq_rendensmat2}
\D_{lm\, l'm',\sigma,\bR} = R_{m\sigma}^{l\dagger}\, 
\Do_{lm\,l'm',\sigma,\bR} \,R^{l'}_{m'\sigma}\,,
\end{align}
with
\begin{align}
R_{m\sigma}^{l} = \begin{cases}
R_{m\sigma}\,;& l=2 \\ 
1\,;& l=1,0
\end{cases}\,,
\end{align}
for the $\bR\neq 0$ part, and
\begin{align}\label{Eq_rendensmat3}
\D_{lm\,l'm',\sigma,\bR=0} = \begin{cases}
\Do_{lm\,l'm',\sigma,\bR=0}\,;& l\neq2,\, l'\neq2\\
\delta_{ll'} n_{m\sigma} & l=2\,,
\end{cases}
\end{align}
with $n_{m\sigma}  = \Tr{\hPhigd \,\hat{n}_{mm',\sigma}\Phig}$,
for the on-site $\bR=0$ part. 
In what follows the matrices $\D$ and $n$ without angular momentum indices will refer to the density matrices of the subset of orbitals with $l=2$. For any value of $l$, we will assume the indices $m$ and $m'$ to run on the cubic harmonics for that value of $l$, which ensures the $\bR=0$ one-body density matrix 
to be diagonal, or, in other words, ``natural''.

The Siesta code provides also the possibility to use a double set of $d$-type orbitals together 
with one set of $s$ and one set of $p$. The use of two sets of $d$-orbital (double-$\zeta$ basis set), 
as opposed to a single set (single-$\zeta$) is particularly indicated for \ac{gga} calculations, 
in which small changes in the density profile of electrons lying close to the Fermi energy can 
affect the calculation of the energy much more than in \ac{lda}. While the first $d$-type basis set 
is more atomic like and more suited for an \ac{ldau} or \ac{ldag} calculation, the second $d$-type set 
has a larger spread, since it is meant to describe better also the tails of the density distribution. For 
\ac{ldag} calculations we will therefore adopt a single-$\zeta$ basis set, while we will use a double-$\zeta$ 
basis set for all \ac{gga} calculations we perform and compare our \ac{ldag} results with.

\section{Results}\label{Sec:results}

\subsection{Nonmagnetic iron}\label{Sec:unpol_orbsel}
\begin{figure}
\includegraphics[width=.8\linewidth]{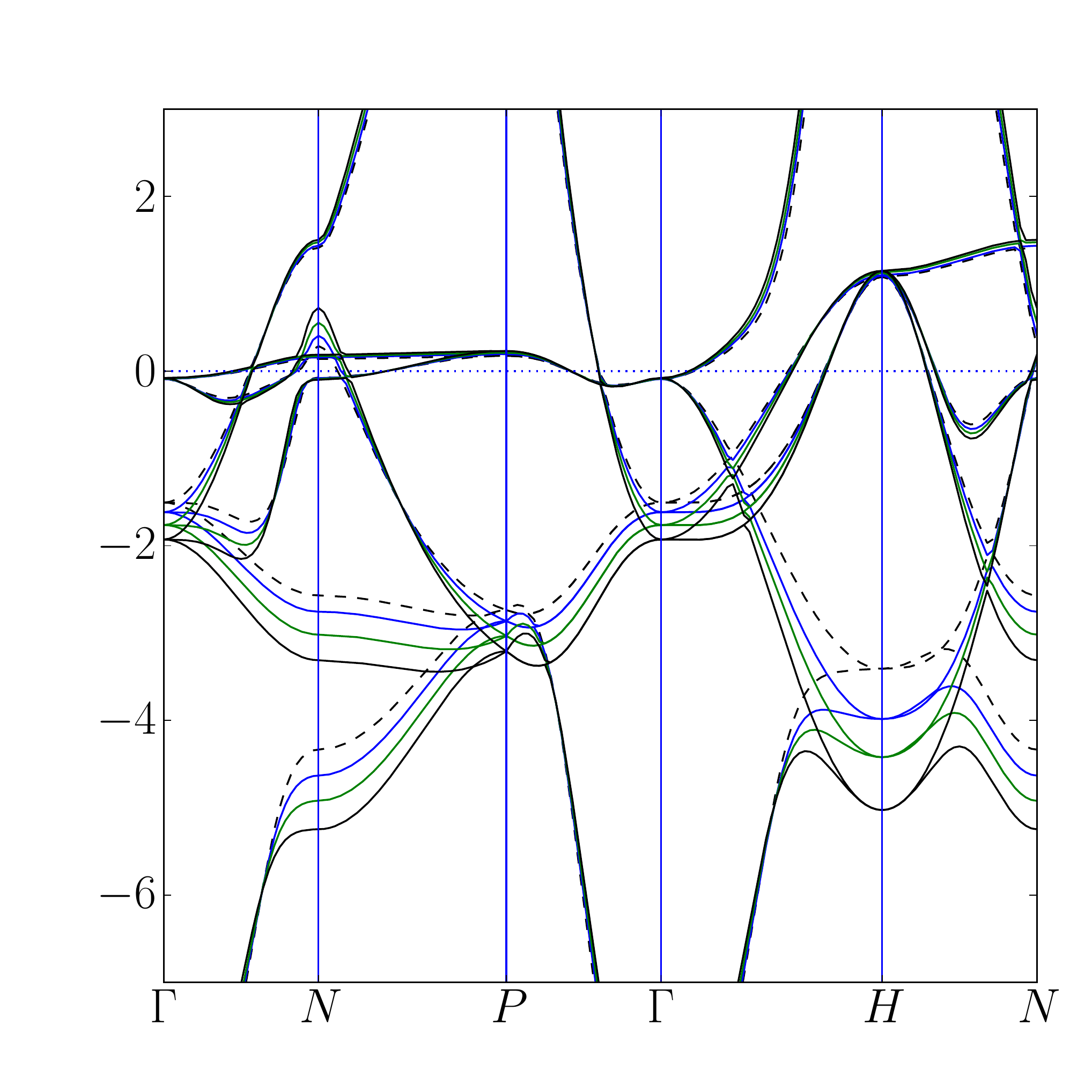}
\caption{(Color online) Band structure results for a Siesta \ac{ldag} calculation of nonmagnetic  $bcc$ iron without optimization of 
the \xno\ $\no_{m\sigma}$. The atomic interaction Hamiltonian we used is displayed in \eqn{Eq:minimodel_hund}, 
and the values for its parameters are listed in \rtab{Tab:fixdensmat_LDA}. The black solid line corresponds 
to $U=0$ and $J=0$, the green line to $U=5$, $J=1.2$, the blue line to $U=10$, $J=1.2$, the dashed line 
to $U=10$ and $J=2.2$. Values of $U$, $J$ and $y$-axis energies are in eV. The labels indicating the 
high-symmetry points for the k-point path are taken from reference~\cite{Setyawan2010299} for the body-centered 
cubic lattice.}\label{Fig:fixdensmat_LDA}
\end{figure}

In order to assess the effect of Gutzwiller renormalization parameters $\Phig$ on the eigenvalues 
of the single-particle Kohn-Sham Hamiltonian at density self-consistency, we show in \fig{Fig:fixdensmat_LDA} 
the band structure of nonmagnetic iron for different values of interaction parameters $U$ and $J$ in the atomic Hamiltonian
\begin{align}\label{Eq:minimodel_hund}
\hHat = U/2 \hN(\hN-\hident) - J|\Spinsq| -\kappa |\Elsq| \,,
\end{align}
where the parameter $\kappa$ has been added in order to 
single out the effect of Hund's third rule. The value of $\kappa\approx0.2$~eV can be estimated from 
spectroscopic data such as those of Corliss and Sugar\cite{Corliss_Sugar};  a guess for a reasonable value 
of $J=1.2$~eV may be obtained either from spectroscopic data or from its expression in terms of Slater 
integrals $F_2$ and $F_4$. Calculation of $F_2$ and $F_4$ has been done using the electronic structure 
program by Cowan\cite{Cowan}, and its results are in agreement with spectroscopic data.
The band structures plotted in \fig{Fig:fixdensmat_LDA} are obtained by performing only the first and 
second optimization steps explained above, 
while the matrix $\nnat_{m\sigma}$ is kept fixed to its \ac{lda} value. In section~\sect{Sec:result_magnetic} 
we will show that even when $\nnat_{m\sigma}$ is treated as a variational parameter, its change with respect to the \ac{lda} value is very small in the case of nonmagnetic iron.
An immediate consequence of the fact that we fix $\nnat_{m\sigma}$ is that we do not need to worry about 
the explicit form of the double-counting energy $\Edc$ for the Hamiltonian \eqref{Eq:minimodel_hund}, 
which plays a role in determining the electronic structure only through the optimization of the natural density matrix. 
In  \rtab{Tab:fixdensmat_LDA} we show the band mass renormalization factors $Z_{\eg}$ and $Z_{\tdg}$ 
for different values of Hubbard parameter $U$, whereas $J$ is kept fixed along all rows of the table but the last one, where it is increased to $2.2$~eV.
Our method, as expected, 
appears unable to reproduce the orbital selective Mott transition of the $\eg$ orbitals, 
obtained by Anisimov \acwk~\cite{anisimov} as a result of a DMFT calculation. In fact, we find only a minor localization of both $\eg$ and $\tdg$ orbitals, driven both by the Hubbard 
interaction $U$ and by Hund's exchange $J$.  The latter 
plays a major role in the orbital-selectivity of the mass enhancement, as can be seen from the last row of 
\rtab{Tab:fixdensmat_LDA}. It is at this stage not possible to clarify how much the weaker orbital selectivity 
resulting from our calculation could be due to the limitations of the Gutzwiller method or perhaps by the fact that our calculation is performed at zero temperature, as opposed 
to the finite temperature approach by Anisimov \acwk. 

\begin{table}
\begin{tabular}{|c|c|c|c|c|c|}
\hline
$U$~(eV) & $\langle(\Delta N)^2\rangle$ & $|S|$  & $|L|$ & $Z_{\eg}$ & $Z_{\tdg}$\\
\hline
 $0^{.}$&2.30&0.89 &3.22 &1. &1. \\
\hline
2.5&1.37&1.00&3.27&0.94&0.96\\
\textcolor{darkgreen}{5} &1.10& 1.03& 3.29&0.90 &0.93 \\
\textcolor{blue}{10} &0.82& 1.04&3.31 &0.82&0.87\\
$10^\ast$ &0.78& 1.25 & 3.05 &0.72&0.82\\
\hline
\end{tabular}
\caption{ Variance of $d$-electron number operator, total spin and angular momentum for $d$ orbitals, 
band mass renormalization factors for $\eg$ and $\tdg$ orbitals for a \ac{ldag} calculation without optimization 
of the natural density matrix, for the atomic Hamiltonian displayed in \eqn{Eq:minimodel_hund}. The value of  
$\kappa$ is $0.2$~eV and the value of $J$ we used was always $1.2$~eV, except for the row marked with 
a dot ($.$), for which $J=0$, and the row marked with an asterisk ($\ast$), for which $J=2.2$. The values of $U$ 
we used are listed in the first column. The band structure results corresponding to the first and to the 
last three rows of the table are plotted in \fig{Fig:fixdensmat_LDA}. The last line of the table shows how 
orbital selectivity is more sensitive to Hund's exchange $J$ than to Hubbard $U$.}\label{Tab:fixdensmat_LDA}
\end{table}

\subsubsection{Impact of hybridization between atomic orbitals on the description of a Mott phase}
The minor correlation-induced enhancement of $\eg$ band mass with respect to the LDA results may be 
connected to the %fact that a 
sizable hybridization connecting $\eg$ orbitals on a site to $s$-orbitals 
on neighboring sites. This hybridization is ineffective close to the 
$\Gamma$ point, where the $e_g$ band remains quite flat, but is able to 
induce an appreciable dispersion in the rest of the Brillouin zone, especially 
close to the $H$ point. The local Gutzwiller projector that we use 
can only provide a $\bk$-independent renormalization $Z$, and is thus unable to distinguish between 
the flat dispersion near the $\Gamma$ point, and e.g., the wider one around the $H$ point. In other words 
$Z$ %has to be better 
should be regarded as an average of the quasi-particle weight over the whole Brillouin zone, indeed a major limitation of the \ac{ga}. 
In the specific example of  $bcc$ iron, it is just the hybridization with the weakly correlated $s$-orbitals that  
prevents a genuine Mott localization of $\eg$ orbitals within the Gutzwiller approximation. In fact, the condition 
for a Mott transition to occur within the Gutzwiller approximation is that the band-energy gain $\Ekin$ upon 
delocalizing quasiparticles does not compensate anymore the cost in 
Hubbard repulsion $\EHub$. In the single-band Hubbard model it can be shown that the Hubbard repulsion 
\[
\EHub = \frac{U}{4}\big(1-\sqrt{1-R^2}\big),
\]
while $\Ekin = R^2\,\Ekin^{(0)}$, where $\Ekin^{(0)}<0$ is the non interacting value. As a result, 
for $U\geq -8\Ekin^{(0)}$ the hopping energy gain cannot compensate anymore the Hubbard repulsion and 
the lowest energy solution is characterized by $R=0$, which describes the Mott insulator. 
When however, as in the present multi-band case, the main part of the band energy is renormalized 
by a {\it single} $R$, we have that $ \Ekin\propto - R$ while still $\EHub\propto R^2$ for $R\ll 1$, 
so that there is always a minimum with $R$ finite, which is what we do find. 

However, in spite of the fact that correlation effects seem not to play a major role in the band structure, 
they substantially affect the magnetic properties, as we are going to show in what follows.   

\subsection{Ferromagnetic and paramagnetic iron: correlation-induced enhancement of local magnetic moment}\label{Sec:result_magnetic}

In order to study magnetic properties of iron, we performed unpolarized and polarized \ac{lda}, \ac{gga} 
and \ac{ldag} calculations, including in the latter also the optimization with respect to the natural density-matrix. 
In Tables~\ref{Tab:dens_zetas} to~\ref{Tab:delta_energies} we list the electronic structure data of $bcc$ iron with 
optimized $\nnat_{m\sigma}$. The adopted values of $U$ and $J$, see 
Eqs. (\ref{Eq:hHat_approssimata}) and (\ref{double-counting-Hund}),    
are $2.5$~eV and $1.2$~eV respectively, both slightly larger than those used by Anisimov \acwk~\cite{anisimov}. 
We observe, see second column of~\rtab{Tab:dens_zetas}, that the optimization 
of $\nnat_{m\sigma}$ in the \ac{ldag} unpolarized case causes only small changes in the matrix elements 
of the natural density matrix with respect to the \ac{lda} result.
This is an {\it a posteriori} justification of the results obtained in \sect{Sec:unpol_orbsel}, and suggests 
that such a value is mainly determined by electrostatic balance,  which is well captured by \ac{lda} 
and does not require a better account of correlation effects. The Gutzwiller parameters do provide the 
wavefunction with more flexibility, but do not seem to give any important feedback on the natural density matrix. 

This feedback becomes instead important in the spin polarized case, where it contributes to an increase 
in total magnetization $m$ as well as of the lattice parameter, as can be seen by comparing the values in 
the second column of \rtab{Tab:lat_spin_energies}.
\begin{table}
\begin{center}
\begin{tabular}{|c|c|c|}
\hline
  & $\no_{\alpha}$ &$m/m^\ast$\\
\hline
 LDA unp. & \scriptsize{0.597,0.685} &\scriptsize{1.,1.}\\
\hline
 LDA pol. & \scriptsize{0.920,0.823,0.303,0.515} &\scriptsize{1.,1.,1.,1.}\\
\hline
 LDA+G unp. & \scriptsize{0.599,0.673}& \scriptsize{0.925,0.953}\\
\hline
 LDA+G pol. & \scriptsize{0.936,0.880,0.277,0.457} & \scriptsize{0.969,0.967,0.984,0.984}\\
\hline
\end{tabular}
\caption{Orbital densities $\no_\alpha$ and quasi-particle mass renormalization $m/m^\ast = R^2_{\alpha\alpha}$ for the different types of simulations performed, with $\alpha=\eg,\tdg$ and $\eg\up,\tdg\up,\eg\dw,\tdg\dw$ for unpolarized (unp.) and polarized (pol.) calculations respectively.}\label{Tab:dens_zetas}
\end{center}
\end{table}

\begin{table}
\begin{center}
\begin{tabular}{|c|c|c|c|c|}
\hline
  & $a_{\rm lat}$ (\AA) & $\;m\;$& $\;m_d\;$ & $2|S|$\\
\hline
 GGA unp. & 2.80 & --- & --- & ---\\
\hline
 GGA pol. & 2.87& 2.33 & --- & --- \\
\hline
 LDA unp. & 2.77 & --- & --- & 1.77\\
\hline
 LDA pol. & 2.83 & 2.066 & 2.14 & 2.61\\
\hline
 LDA+G unp. & 2.86 & --- & --- & 2.47 \\
\hline
 LDA+G pol. & 2.87 & 2.44 & 2.58& 3.04 \\
\hline
 Exp. & 2.87 & 2.22 & --- &  --- \\
\hline
\end{tabular}
\caption{Results for optimized lattice parameter $a_{\rm lat}$ (in \AA), total magnetization $m$, magnetization 
$m_d$ on $d$-type orbitals, and total spin $2|S|$ (in Bohr magnetons) on $d$ orbitals. 
The last row shows the experimental values for lattice parameter and magnetization.}\label{Tab:lat_spin_energies}
\end{center}
\end{table}

Within our Gutzwiller approach we are also able to compute the local spin moment $|S|$ on $d$-type orbitals, from the expectation value of $\Spinsq$
\begin{align}\label{Eq:spinsq_Gpar}
\tspinsq(\tspinsq+1) = \Tr{\hPhigd\Spinsq\hPhig}\,.
\end{align}
This magnetic moment is partially aligned to the $z$ axis, thus contributing to the total magnetization $m$, 
which is instead computed from the Gutzwiller-renormalized density $n(\br)$ as
\begin{align}
m = \int d\br \left[\nup(\br)-\ndw(\br)\right]\,.
\end{align}

It is worth remarking that the increase of $2|S|$ from \ac{lda} to \ac{ldag} polarized calculations is almost equal 
to the simultaneous increase in magnetization, suggesting that the magnetization rise caused by \ac{ldag} is mainly 
due to the larger local magnetic moment available due to correlations.

\begin{figure*}
\begin{minipage}{.45\linewidth}
\includegraphics[width=\linewidth]{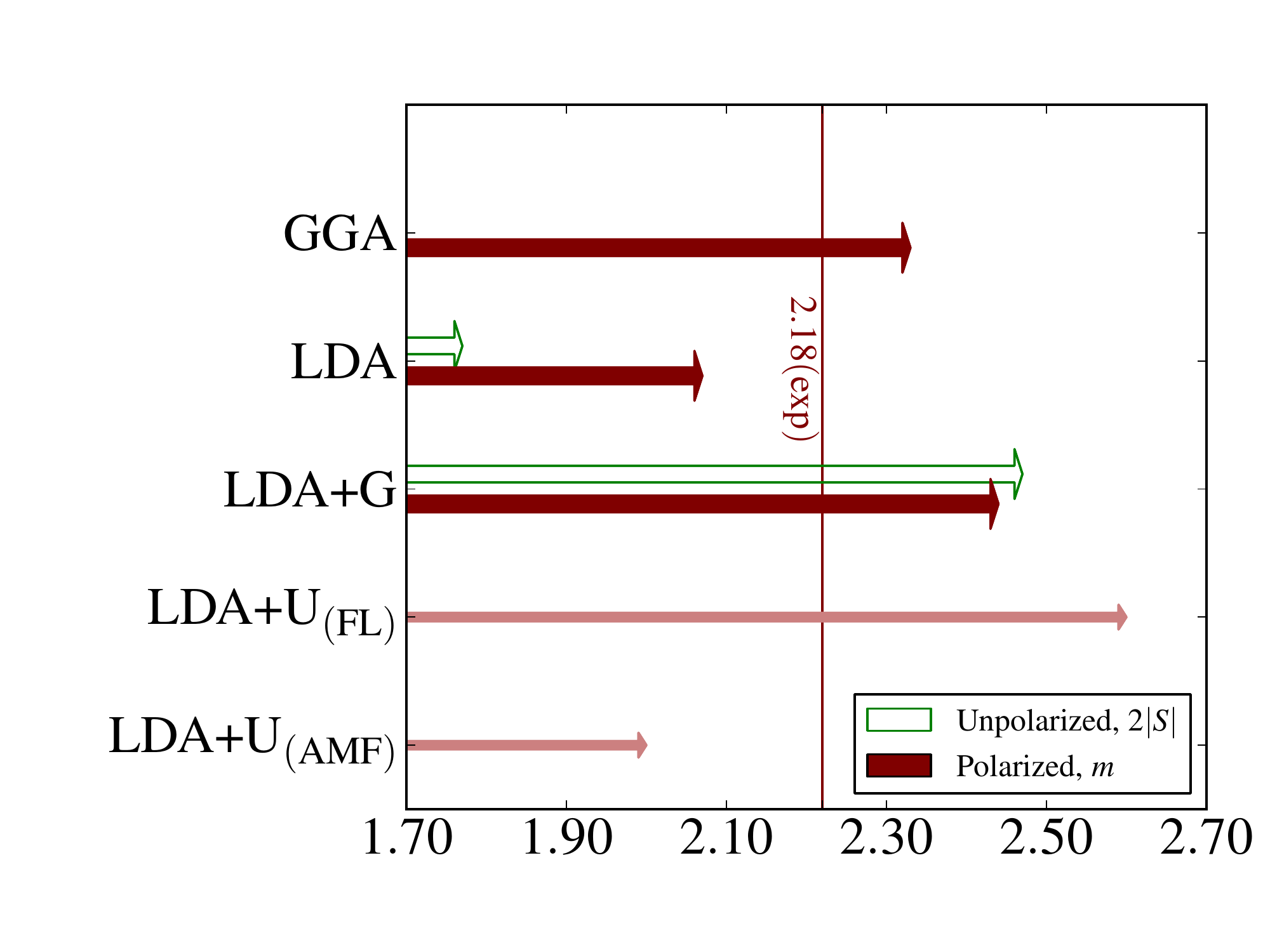}
\end{minipage}
\begin{minipage}{.45\linewidth}
\includegraphics[width=\linewidth]{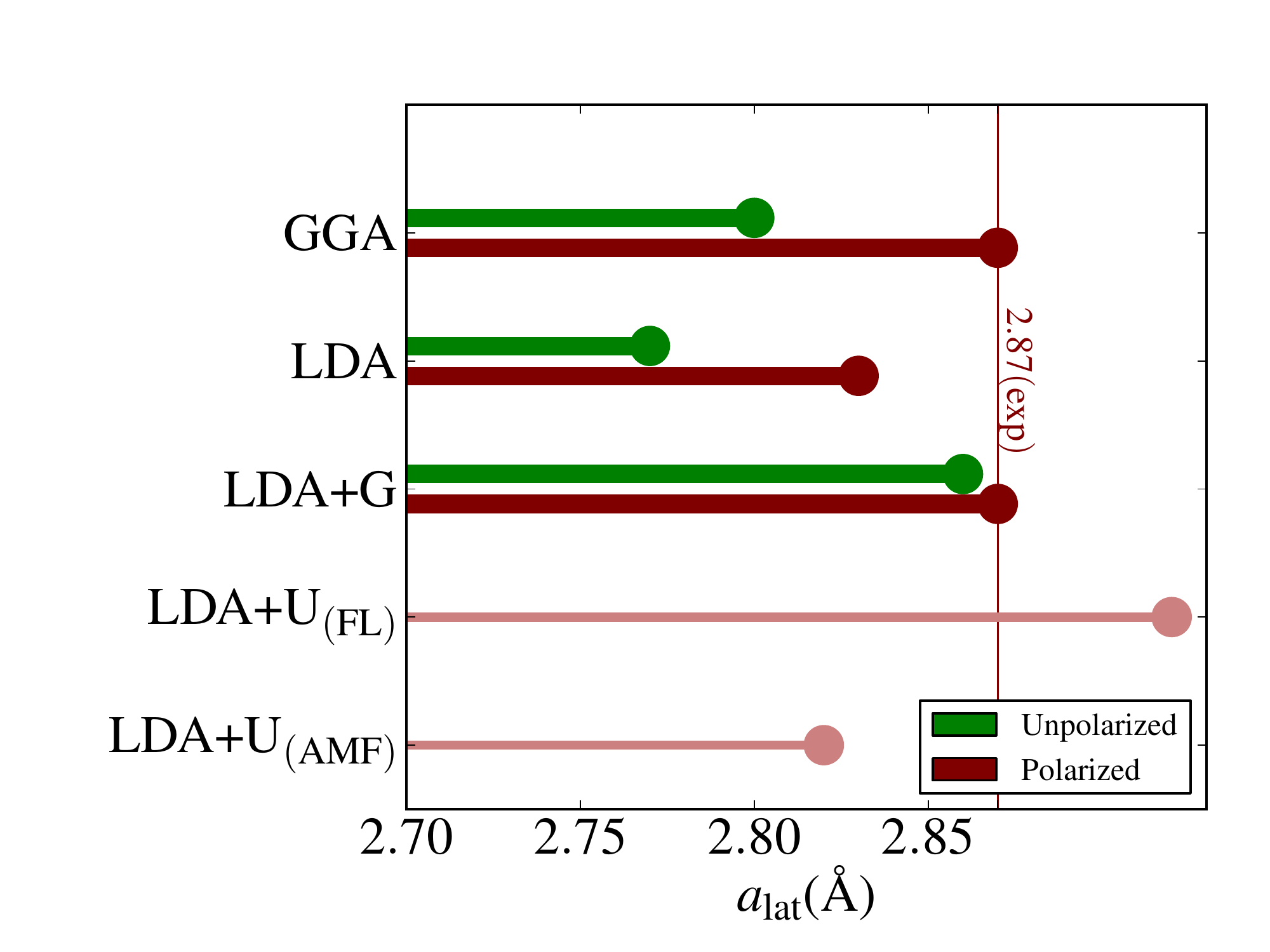}
 \caption{}\label{Fig:alat}
\end{minipage}

\begin{minipage}{\linewidth}
\caption{(Color online) Top left: magnetic moment $2|S|$ (for unpolarized calculations, empty green arrows) and 
magnetization $m$ (for polarized calculations, filled dark red arrows) within GGA, \ac{lda}, \ac{ldag}. 
The last two thin pink arrows refer to previous calculations by Cococcioni and De Gironcoli~\cite{Cococcioni_DeGironcoli_LinLDAU}. 
Top right: lattice parameters listed for the same calculations (green unpolarized, red polarized, pink from reference~\cite{Cococcioni_DeGironcoli_LinLDAU}).
The figures plotted in these graphs are listed also in \rtab{Tab:lat_spin_energies}.}\label{Fig:alat_magneticmoment}
\end{minipage}
\end{figure*}

\subsection{Energy balance in ferromagnetic vs. paramagnetic iron}\label{Sec:balance} 

We observe that, when magnetism is allowed, the magnitude of the $d$-orbital local moment $|S|$ increases 
less within LDA+G, roughly 20\%, than LDA, around 50\%.  In other words, nonmagnetic, unpolarized LDA+G already provides 
iron with local moments of the right order of magnitude, ready to align together when given the possibility.  The
%flair for
propensity towards magnetic order %that we find by 
in LDA+G has its counterpart in the balance of the various contributions 
to the total energy.  In~\rtab{Tab:energies} we list the total 
energies of the various density functional calculations which we carried out, and 
in \rtab{Tab:energies_comparisons} the energy differences between polarized and unpolarized calculations. 
In both tables the total energy is divided up into kinetic, electron-ion interaction plus electrostatic, and exchange-correlation 
contributions. In \rtab{Tab:delta_energies} we indicate the error of LDA+G 
arising from the linearization \eqref{Eq:Hartree1} and~\eqref{Eq:exc1} 
of the Hartree and exchange-correlation energies, respectively. We observe that these errors are 
much smaller than the energy differences in \rtab{Tab:energies_comparisons}, which are therefore reliable estimates.

\begin{figure}
\includegraphics[width=\linewidth]{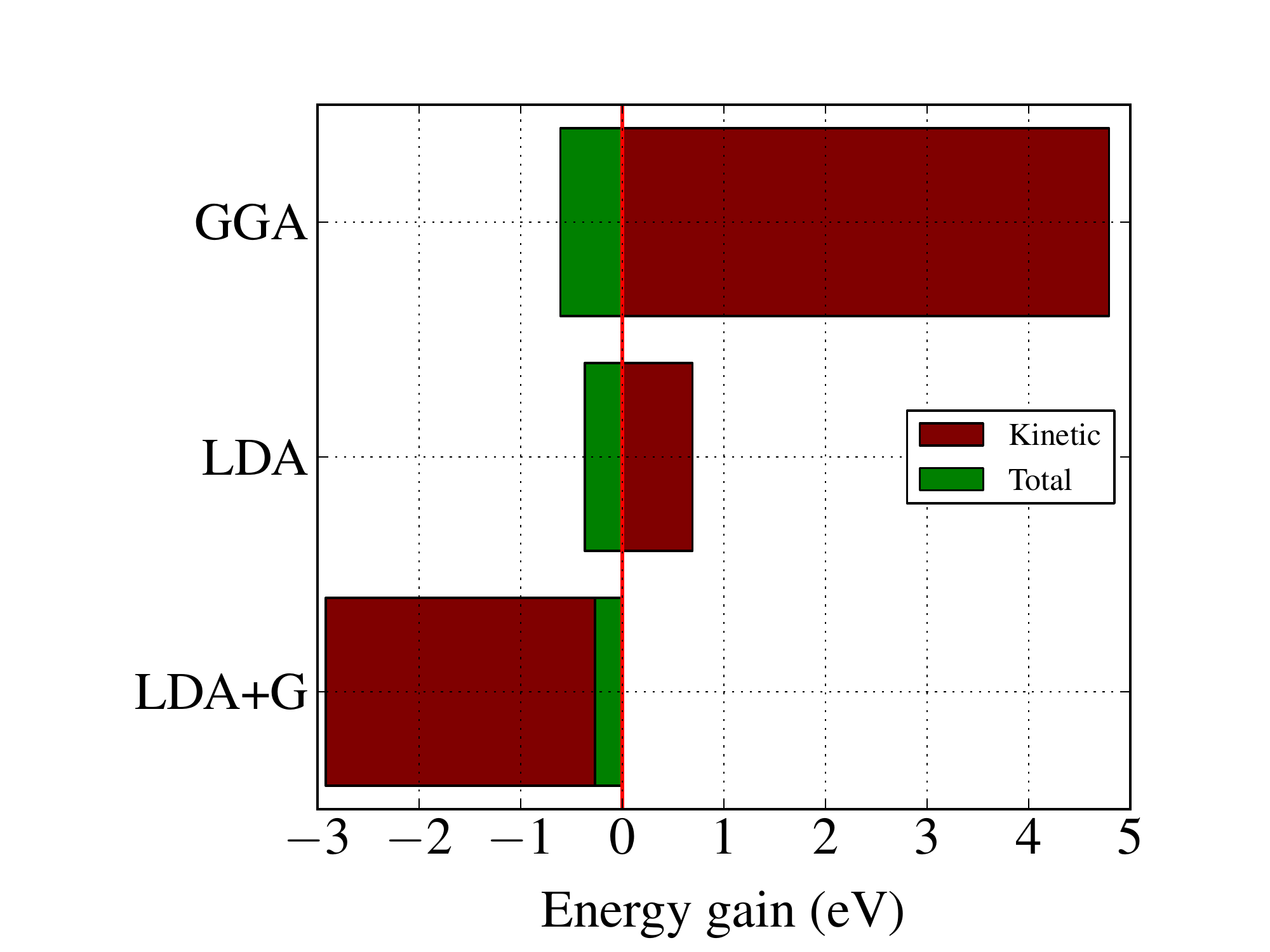}
\caption{(Color online) Total (green) and kinetic (red) energy gain (in eV per atom) of the spin polarized relative to the 
spin unpolarized phase of $bcc$ iron. The values plotted are listed along the second and third columns 
of~\rtab{Tab:delta_energies}. }\label{Fig:energy_diff}
\end{figure}

Focusing on the last two rows in~\rtab{Tab:energies_comparisons}, we note a most interesting  
fact. While in \ac{lda} the onset of magnetism is accompanied by a loss of kinetic energy 
overwhelmed by a gain in electron-ion, Hartree and exchange potential energies, {\it the opposite} actually
occurs in \ac{ldag}. Upon addition of electron correlation effects  through \ac{ldag},
$bcc$ iron emerges as a correlated material, where magnetism appears as the ordering of pre-existing 
moments driven by kinetic rather than potential energy gain. With the moderate and controlled improvement represented by 
Gutzwiller projection over LDA, the basic physical reason why the atomic magnetic moments of iron order 
ferromagnetically turns completely around, from the inter-site exchange upon which the itinerant Stoner picture is based, to one 
that is closer to double exchange. Ferromagnetism is necessary to allow the $\tdg$ electrons (actually "holes")  
to propagate and thus reduce their kinetic energy,\cite{Zener_doublexch,Anderson_doublexch} as in the 
manganites.\cite{manganites}

More in detail, the gain in kinetic energy is signaled by the fact that the quasi-particle weights $Z$ increase
when a finite magnetization is allowed to appear, as can be seen in Table~\ref{Tab:dens_zetas}. This result,
that quasiparticles propagate better in presence of ferromagnetic order, supports the view of a 
double-exchange mechanism, a suggestion that would be better  
evidenced
if we could have a 
$\bk$-resolved
evaluation of quasiparticle weights. Double-exchange is an intrinsically many-body 
effect 
which cannot be reproduced within theories that do not include bandwidth, or kinetic energy renormalization due 
to electron-electron interaction.  For this reason it was not uncovered by previous DFT calculations,
whereas it would naturally occur as a consequence of Anisimov's selective $\eg$ Mott localization.  
The double-exchange process can only set in in presence of 
long-lived on-site magnetic moments, independently of their inter-site ordering.
An independent-electron, single-Slater-determinant theory as \ac{hf}, or \ac{lda} can only describe the 
birth of a magnetic moment  through the simultaneous appearance of a net spin polarization.  This same magnetization, hence  the presence of a majority spin component, is able to decrease electron-electron interaction energy through the Pauli exclusion principle and to increase electron localization 
because of phase-space shrinking. The latter has as a natural effect also the increase of kinetic energy, a characteristic result of the so-called 
Stoner-Wohlfarth picture\cite{Stoner} that we now find in 
disagreement with our results for iron.

\begin{table}
\begin{center}
\begin{tabular}{|c|c|c|c|c|}
\hline
 & $E_{\rm tot}$ & $E_{\rm kin}$ & $E_{\rm at+el}$ & $E_{\rm xc}$\\
\hline
 GGA unp. & -781.625  & 765.108 &-1157.611&-389.121\\
 GGA pol. & -782.235 & 769.901 &-1161.603&-390.533\\
\hline
 LDA unp. & -780.196& 777.255 &-1170.507&-386.943\\
 LDA pol. & -780.567& 777.947 &-1171.205&-387.308\\
\hline
 LDA+G unp. & -777.231 & 777.099 &-1168.651&-385.682\\
 LDA+G pol. & -777.499 & 774.182 &-1165.568&-386.117\\
\hline
\end{tabular}
\caption{Total energy (eV/atom) for $bcc$ iron computed with the different basis sets and functionals, 
divided in total energy, kinetic energy, atomic interaction plus electrostatic energy $E_{\rm at+el}$, and 
exchange-correlation energy. The quantity on the fourth column is equal to 
$E_{\rm at+el} = E_{\rm ion}+E_{\rm ie}+E_{\rm H}+\Eat-\Edc$, where $\Eat$ is defined in 
\eqn{Eq:eat}, $\Edc$ is defined in \eqn{double-counting-Hund}, and $E_{\rm ie}$ and $E_{\rm ion}$ are the electrostatic 
interaction energies between ions and electrons and between ions and ions. The}\label{Tab:energies}
\end{center}
\end{table}

\begin{figure*}
\begin{center}
\begin{minipage}{.45\textwidth}
\includegraphics[width=.75\textwidth]{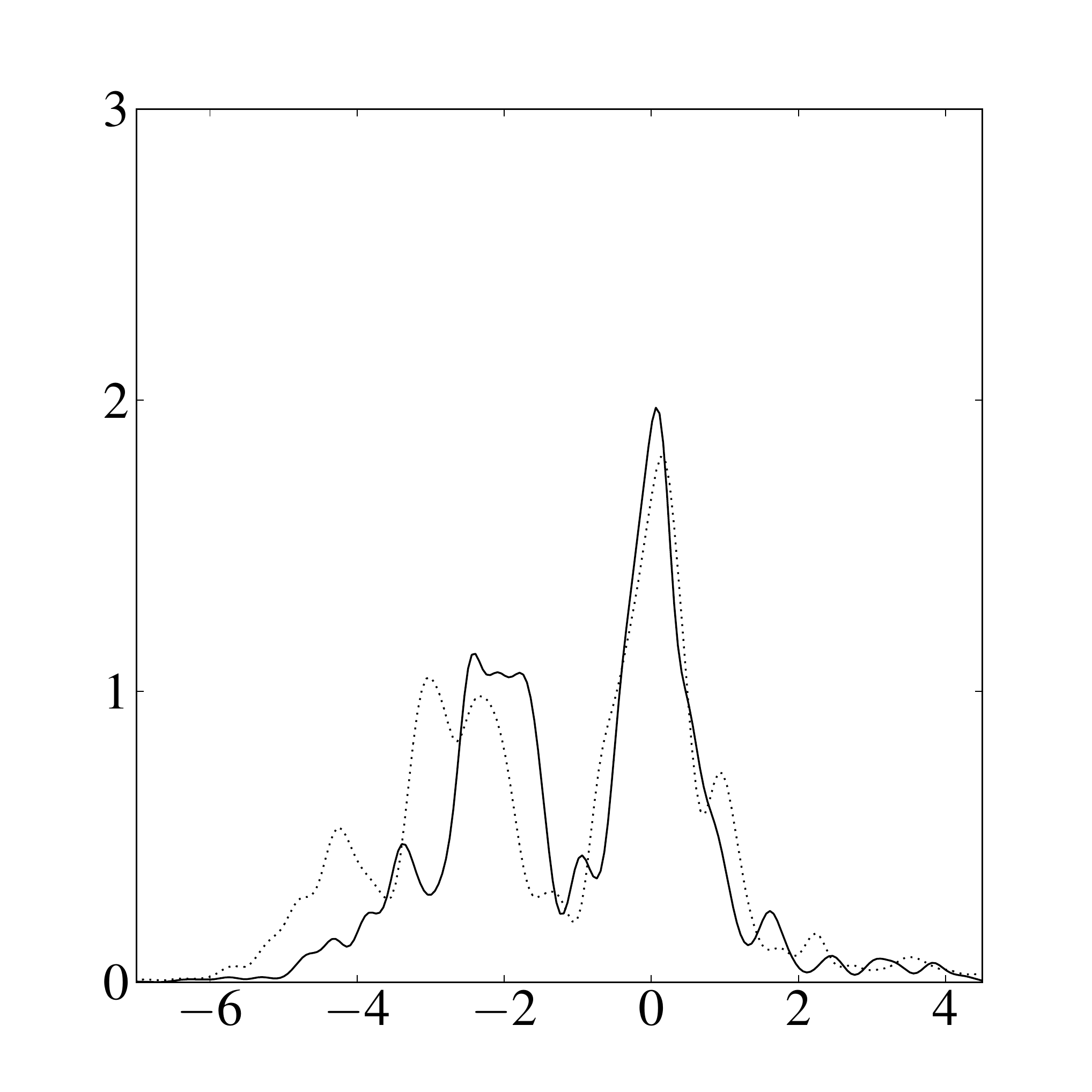}
\end{minipage}
\begin{minipage}{.45\textwidth}
\includegraphics[width=.75\textwidth]{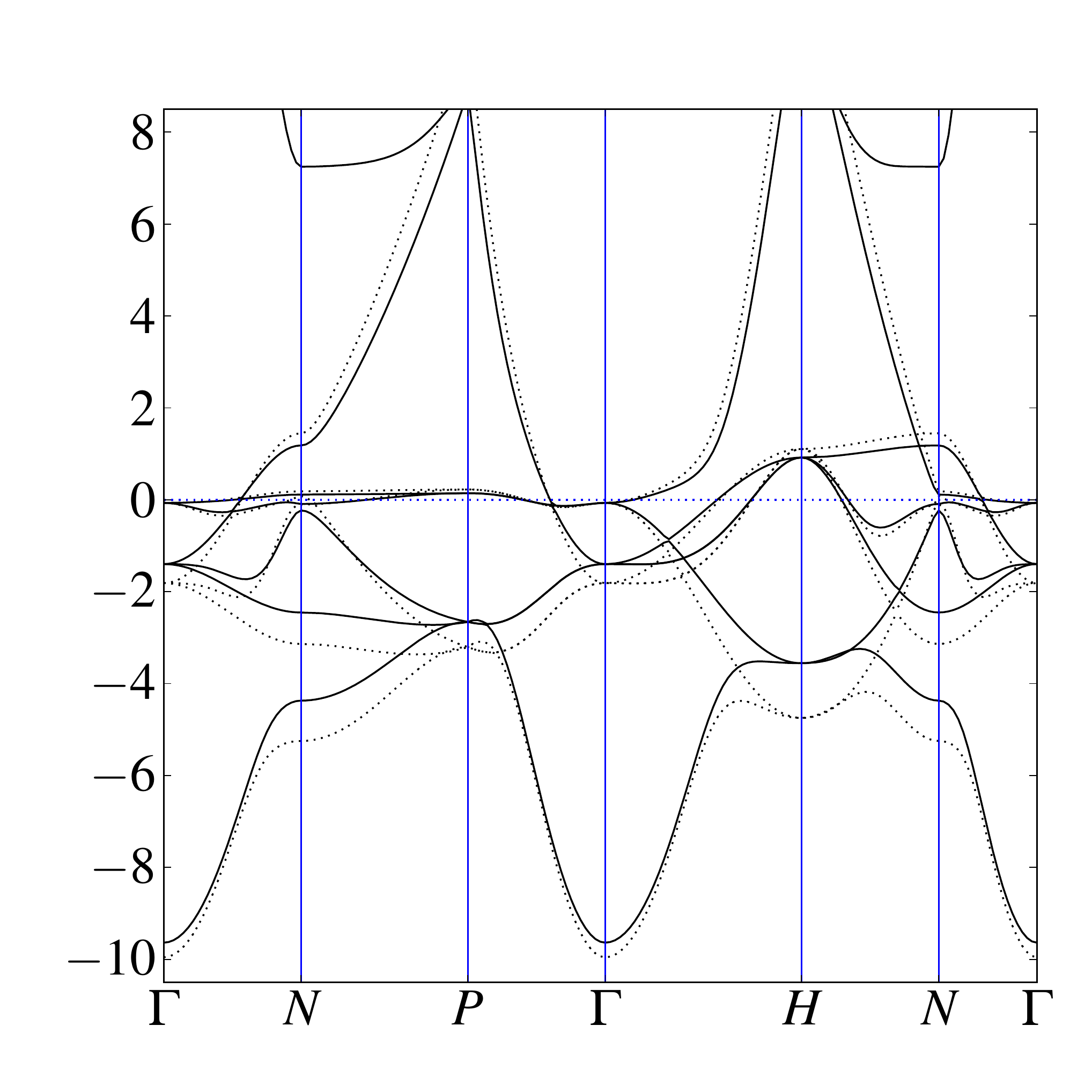}
\end{minipage}

\begin{minipage}{\textwidth}
\caption{(Color online) Comparison of projected density of states and band structure between spin unpolarized \ac{ldag} (solid lines) 
and \ac{lda} (dotted lines).}\label{Fig:para_single-zeta_lda}
\end{minipage}
\end{center}
\end{figure*}

\begin{figure*}
\begin{center}
\begin{minipage}{.45\textwidth}
\includegraphics[width=.75\textwidth]{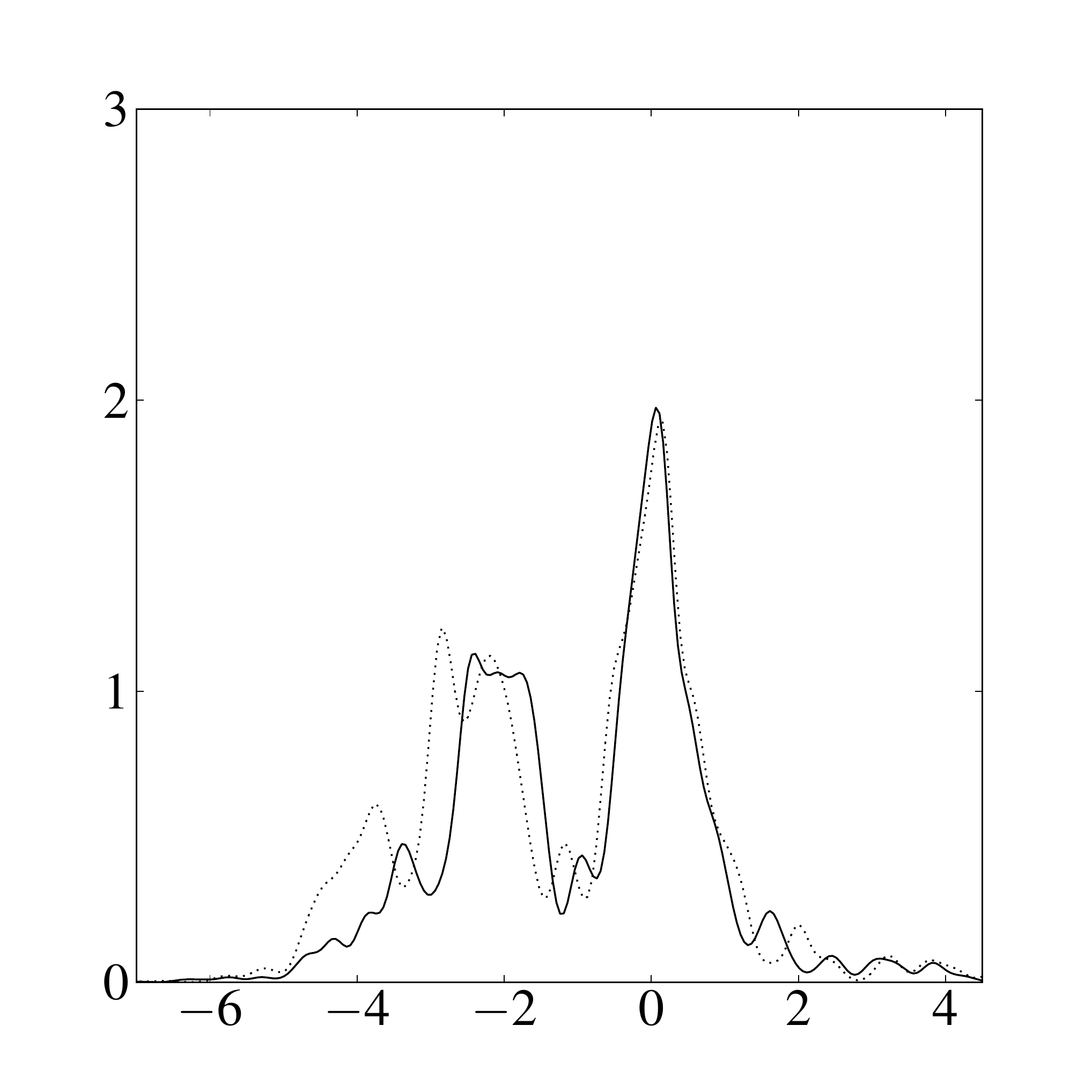}
\end{minipage}
\begin{minipage}{.45\textwidth}
\includegraphics[width=.75\textwidth]{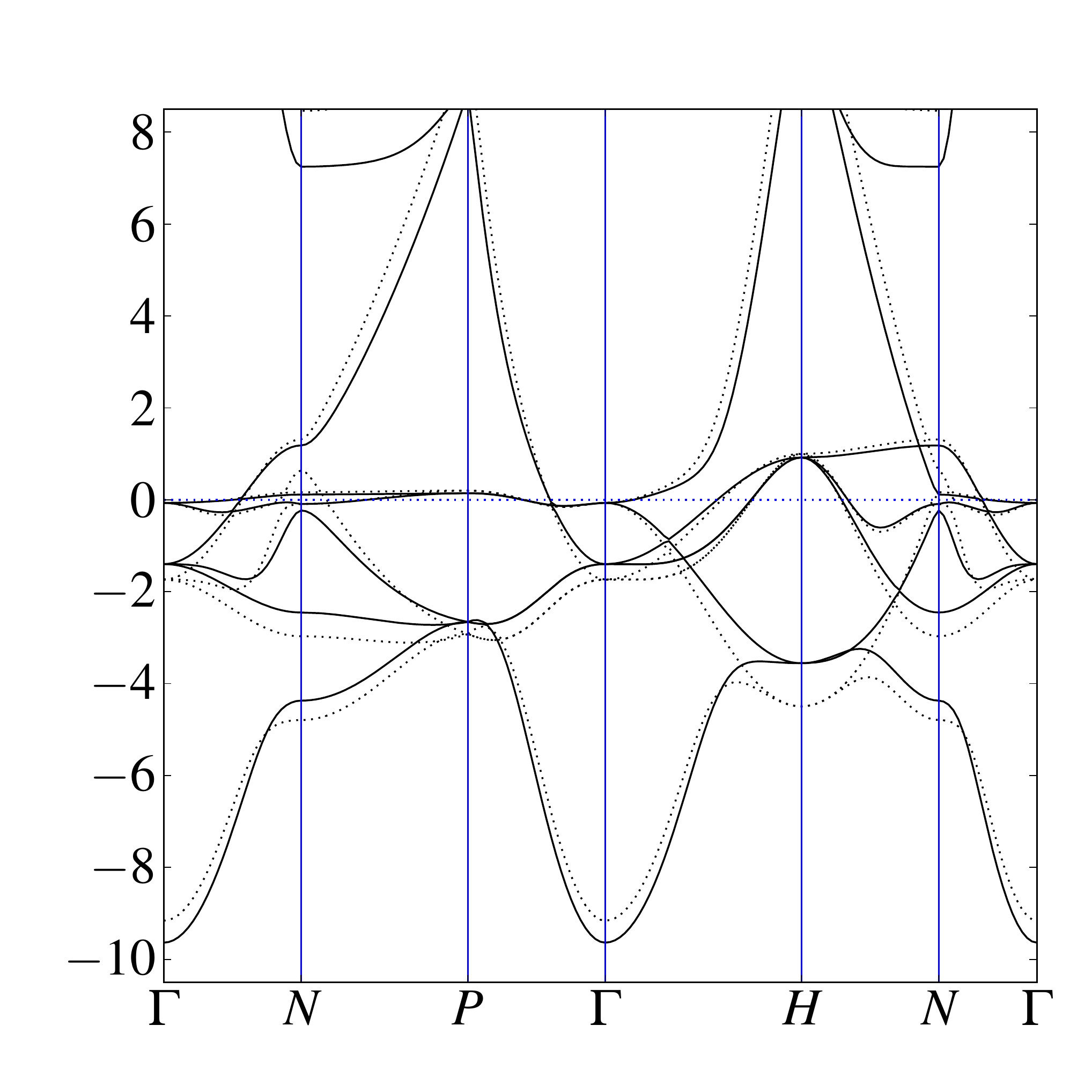}
\end{minipage}

\begin{minipage}{\textwidth}
\caption{(Color online) Comparison of projected density of states and band structure between spin unpolarized \ac{ldag} (solid lines) 
and \ac{gga} (dotted lines).}\label{Fig:para_single-zeta_gga}
\end{minipage}
\end{center}
\end{figure*}

\begin{figure*}
\begin{center}
\begin{minipage}{.30\textwidth}
\includegraphics[width=.75\textwidth]{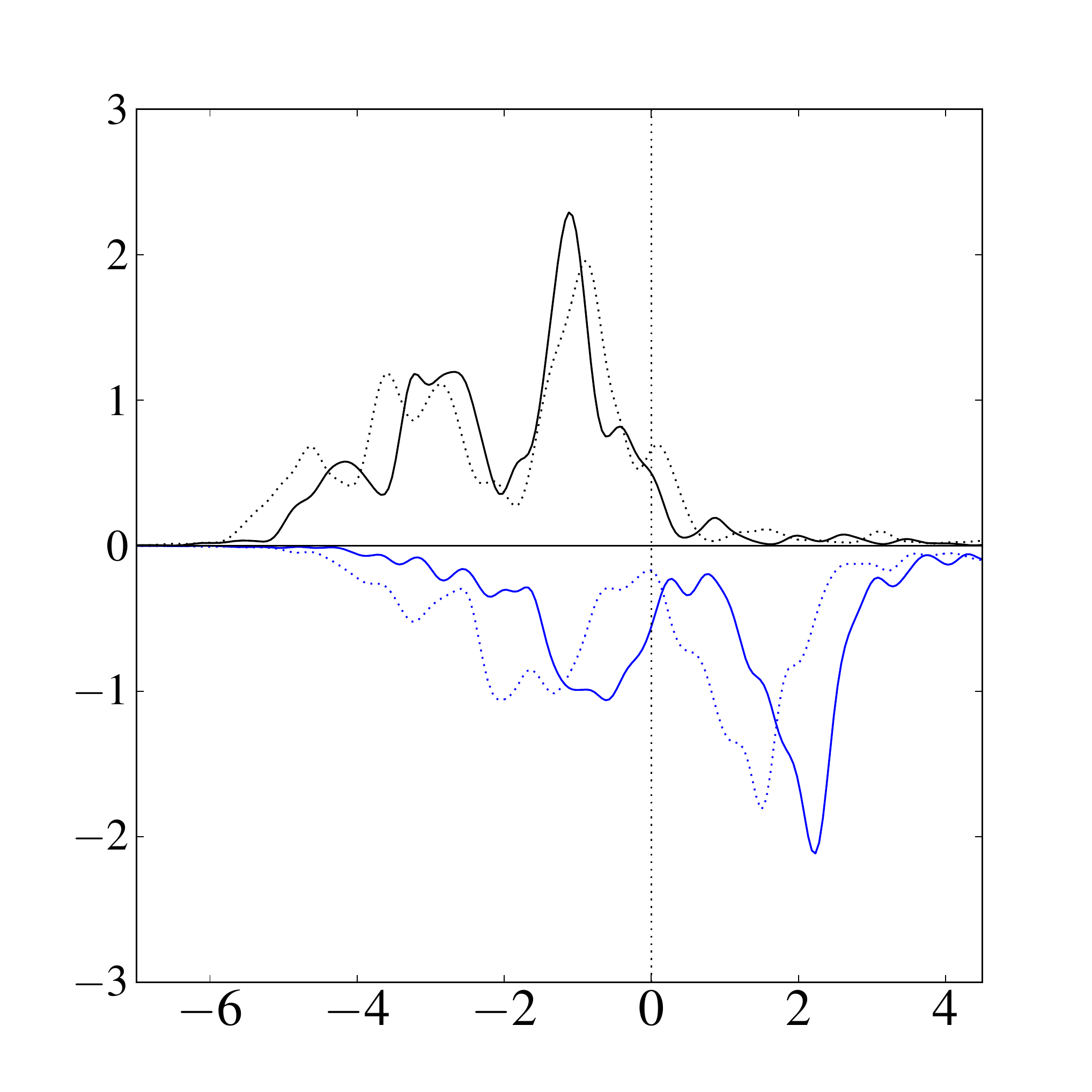}
\end{minipage}
\begin{minipage}{.30\textwidth}
\includegraphics[width=.75\textwidth]{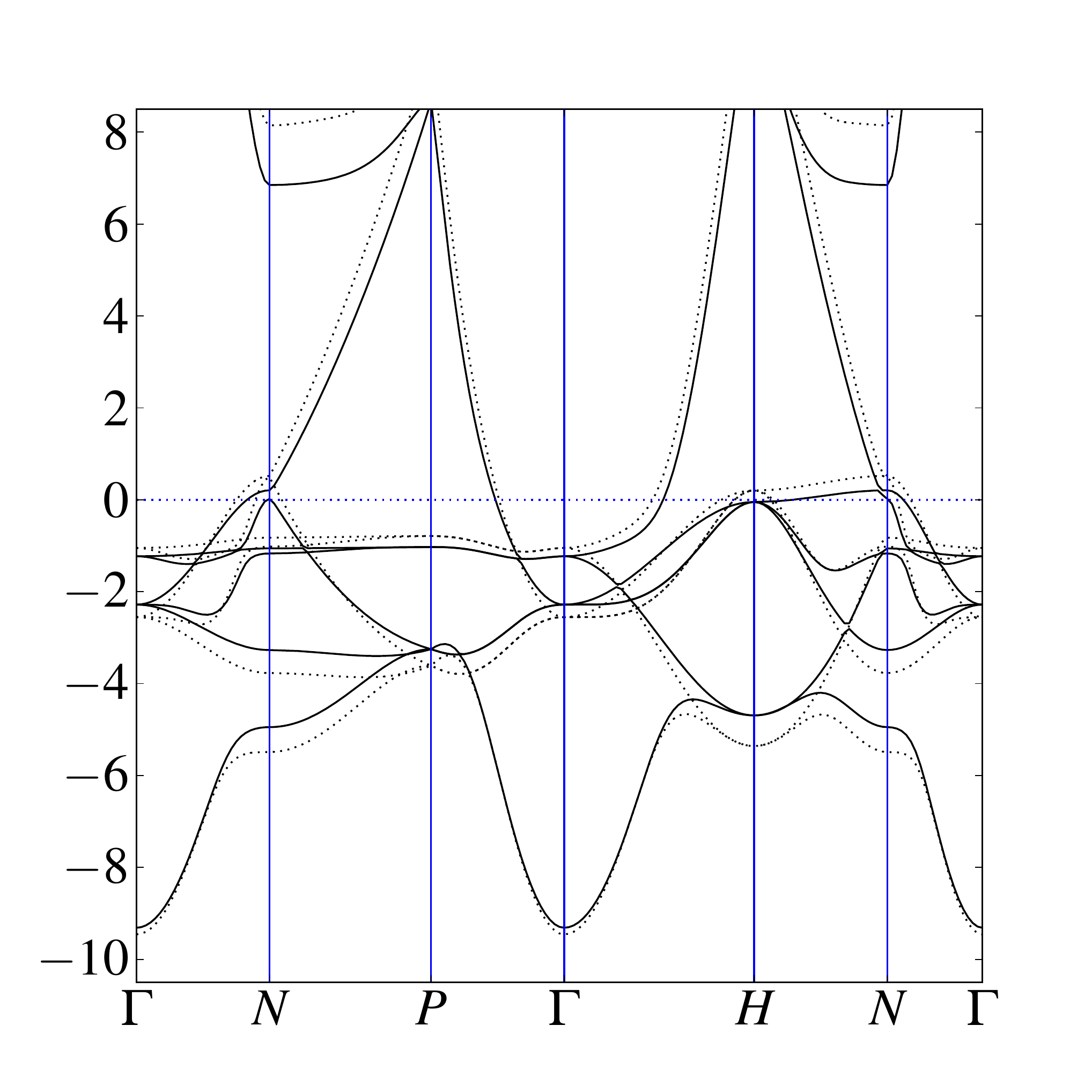}
\end{minipage}
\begin{minipage}{.30\textwidth}
\includegraphics[width=.75\textwidth]{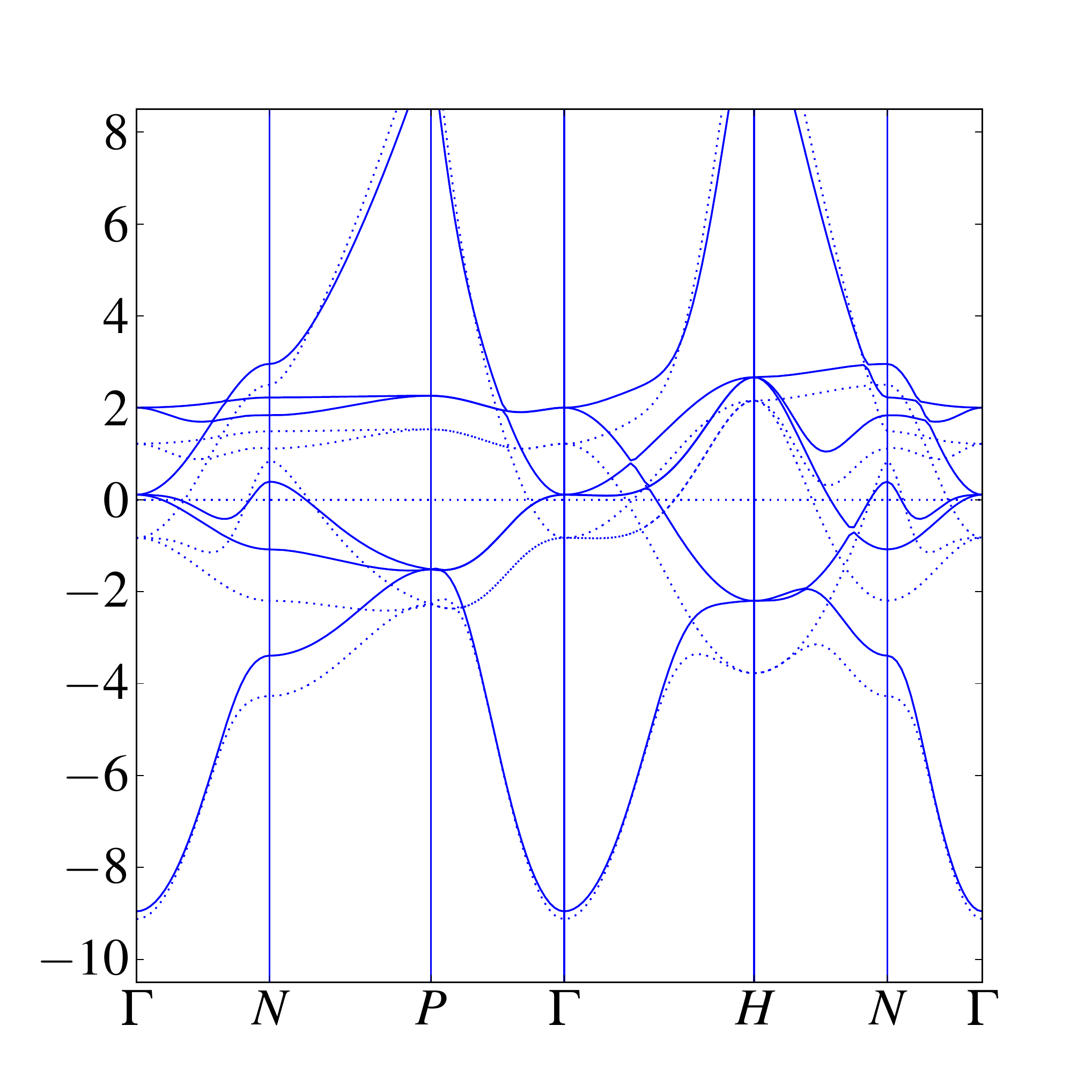}
\end{minipage}

\begin{minipage}{\textwidth}
\caption{(Color online) Comparison of projected density of states and band structure between spin polarized \ac{ldag} (solid lines) and \ac{lda} 
(dotted lines). The line colors blue and black refer to minority and majority component respectively.}\label{Fig:ferro_single-zeta_lda}
\end{minipage}
\end{center}
\end{figure*}

\begin{figure*}
\begin{center}
\begin{minipage}{.30\textwidth }
\includegraphics[width=.75\textwidth]{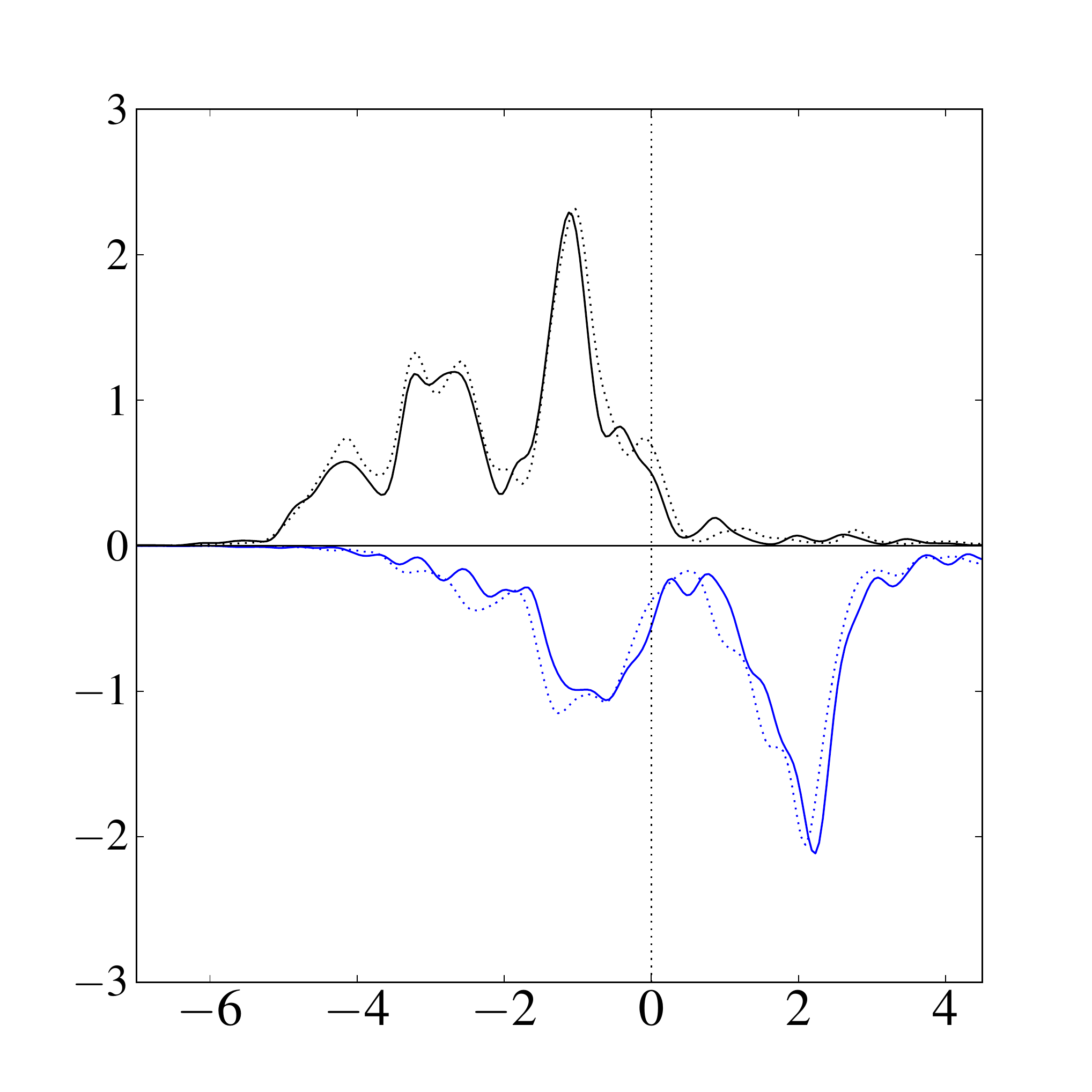}
\end{minipage}
\begin{minipage}{.30\textwidth}
\includegraphics[width=.75\textwidth]{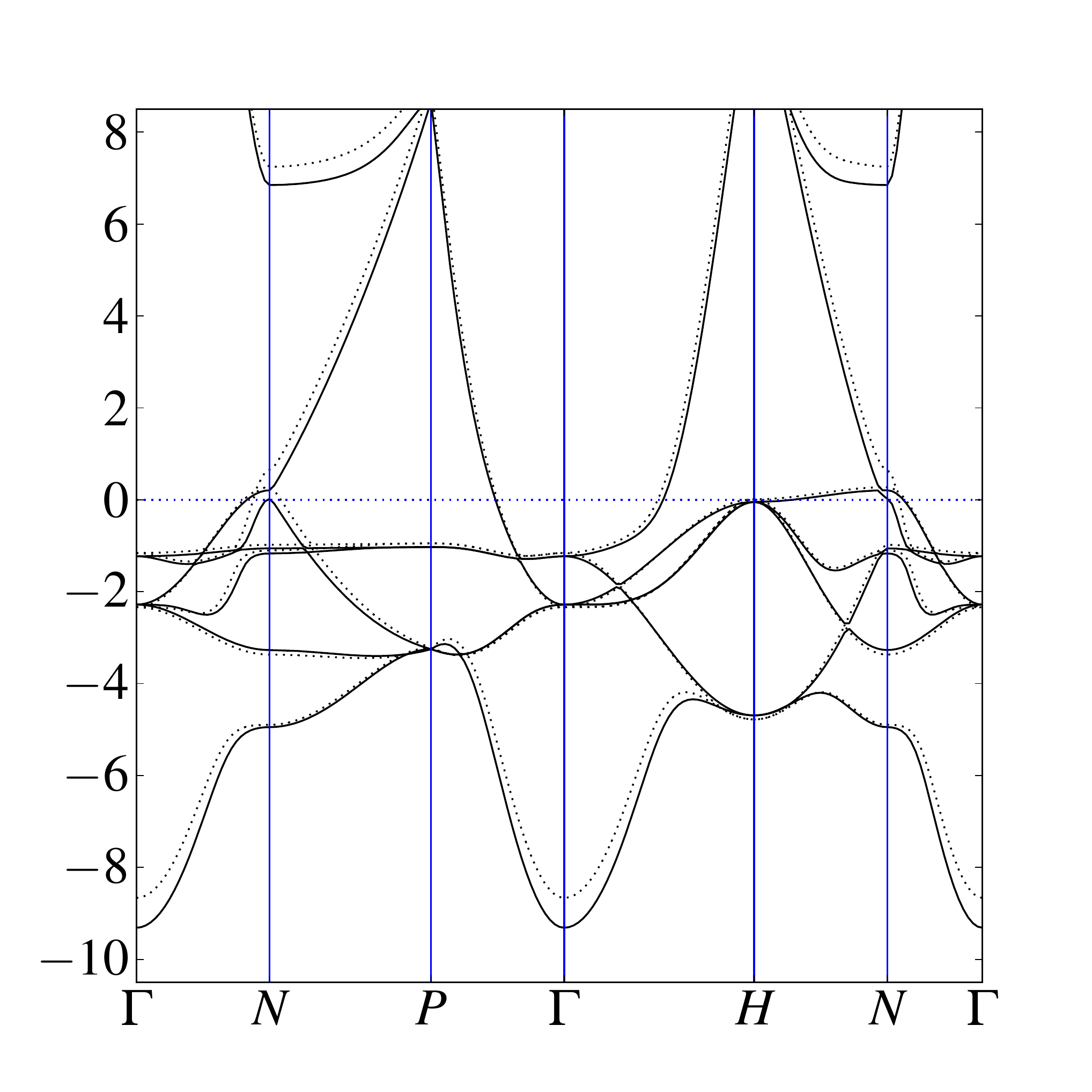}
\end{minipage}
\begin{minipage}{.30\textwidth}
\includegraphics[width=.75\textwidth]{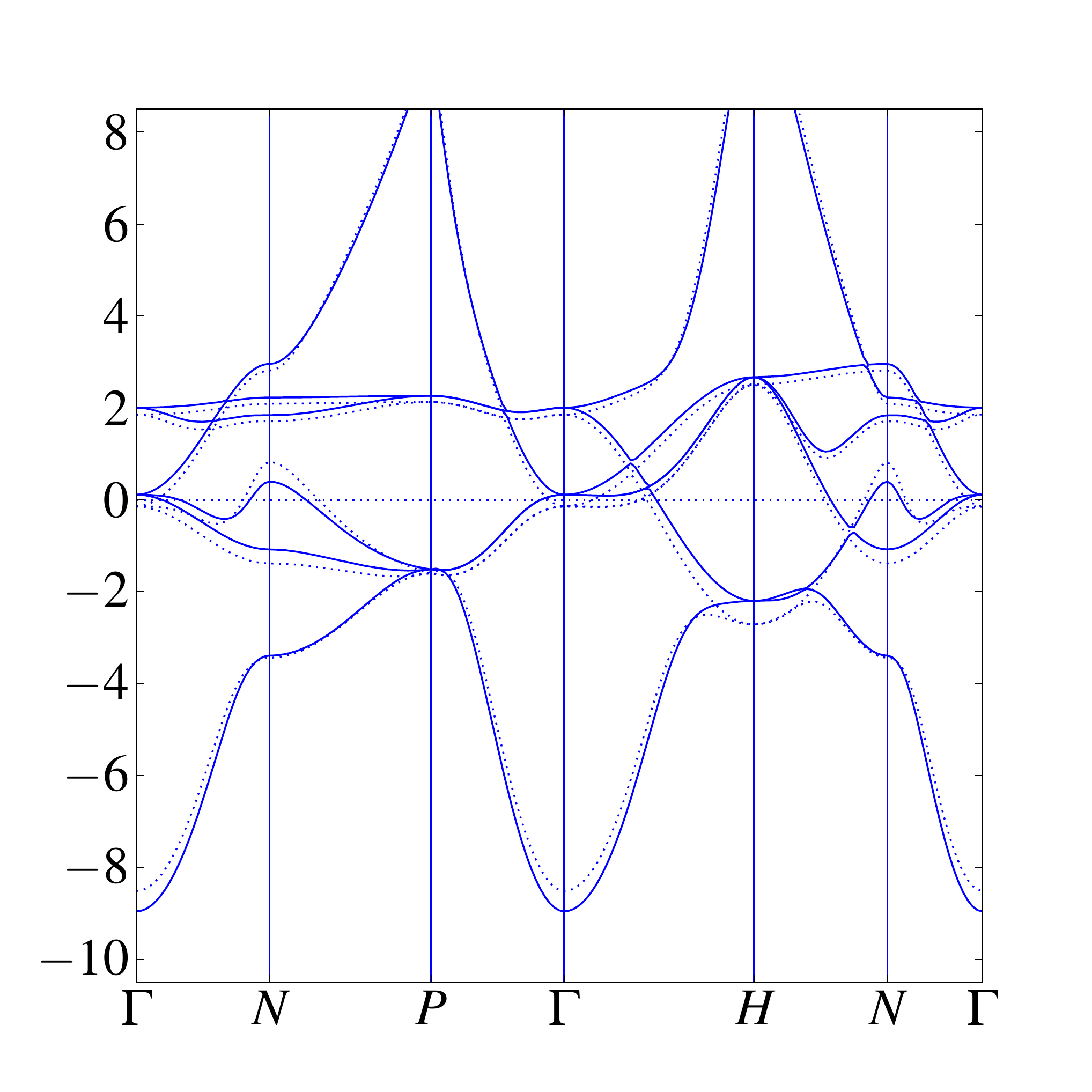}
\end{minipage}

\begin{minipage}{\textwidth}
\caption{(Color online) Comparison of projected density of states and band structure between spin polarized \ac{ldag} (solid lines) 
and \ac{gga} (dotted lines). The line colors blue and black refer to minority and majority component respectively.}\label{Fig:ferro_single-zeta_gga}
\end{minipage}
\end{center}
\end{figure*}

\begin{table}
\begin{center}
\begin{tabular}{|c|c|c|c|c|}
\hline
 & $\delta E_{\rm tot}$&$\delta E_{\rm kin}$&$\delta E_{\rm at+el}$& $\delta E_{\rm xc}$\\
\hline
 GGA & -0.61 &4.79 & -3.992&-1.412 \\
\hline
 LDA & -0.37 & 0.692 & -0.698 &-0.365 \\
\hline
 LDA+G & -0.27 & -2.92 & 3.083 &-0.44\\
\hline
\end{tabular}
\caption{Energy differences (eV/atom) between the spin-polarized and unpolarized ground-states of $bcc$ Fe, 
taken from \rtab{Tab:energies}. By looking at the last two columns, one notices the opposite signs of 
kinetic and $E_{\rm at+el}$ gains when switching between the two ground-states. The kinetic energy gain is
connected with the increase of Gutzwiller band mass renormalization factors $Z$ from spin unpolarized to 
spin polarized wavefunctions, as can be seen from \rtab{Tab:dens_zetas}.
}\label{Tab:energies_comparisons}
\end{center}
\end{table}
The results outlined above, although in qualitative agreement with the views of Anisimov {\sl et al.}, 
are sufficiently strong to call for an independent assessment for the accuracy of the Gutzwiller approach.
A comforting check is provided by comparison of the band structures and density of states obtained within 
\ac{lda} , \ac{ldag} and \ac{gga}, shown from \fig{Fig:para_single-zeta_lda} 
to \fig{Fig:ferro_single-zeta_gga}, and from \fig{Fig:alat_magneticmoment}, showing the estimated 
lattice parameters, magnetization and total magnetic moments within the same three functional schemes.
\begin{table}
\begin{center}
\begin{tabular}{|c|c|c|}
\hline
  & $\Delta E_{\rm H}$&$\Delta E_{\rm xc}$\\
\hline
 LDA+G unp. & 0.0083&-0.0012\\
\hline
 LDA+G pol. & 0.0054 & -0.0020\\
\hline
\end{tabular}
\caption{Estimated errors (in eV) due to the approximate expressions~\eqref{Eq:Hartree1} and~\eqref{Eq:exc1} 
for the Hartree and exchange-correlation energies respectively, listed for the spin unpolarized and spin polarized 
ground-state calculations. All errors are negligible with respect to the energy differences computed in \rtab{Tab:energies_comparisons}.}\label{Tab:delta_energies}
\end{center}
\end{table}
The spin-polarized \ac{gga} is generally considered as a reliable approach to transition metals, as is able 
to provide a very good estimation of their lattice constants and magnetic moments. The Siesta \ac{gga} 
prediction for the iron lattice parameter is 2.87~\AA, in good agreement with the experimental value, while 
its magnetic moment is slightly overestimated (2.33 vs. 2.22 Bohr magnetons, see ). 
We note that \ac{ldag} corrects, without a need for gradient terms, the lattice parameter underestimation which is a well known flaw of \ac{lda}. \ac{ldag} also increases the total 
magnetic moment from the underestimated \ac{lda} value to a slightly overestimated one, now larger than 
the \ac{gga} result (see again \fig{Fig:alat_magneticmoment}).
Comparing the polarized band structures, we note an upward shift of the minority band in LDA+G with respect to 
GGA, consistent with the larger magnetization obtained by the former, although the shapes are quite similar. In particular, the minority band at $\Gamma$ within LDA+G lies above the Fermi energy, 
while it is below in GGA, 
and in experiments. \cite{ARPES-iron} %Indeed t
The 
detailed 
band behavior near $\Gamma$ is notoriously delicate, as recently discussed by Ref. ~\onlinecite{Philipp_Werner}, 
and crucially depends on all parameters that contribute to determine the precise value of magnetization, not 
least the uncertainty in the expression of the double-counting term.    
Previous \ac{ldau}  calculations on iron\cite{Cococcioni_DeGironcoli_LinLDAU, mazin} indeed pointed 
out the differences arising by using an around-mean-field
instead of a fully-localized expression for the double-counting 
energy, a question that  would be worth of further investigation.\cite{Sasha-DC} 

\section{Final remarks}

We presented here an explicit and careful  
implementation of a 
self-consistent scheme integrating the standard local density functional formalism of Kohn and Sham \ac{dft} with that 
implied by Gutzwiller wavefunction corrections\ac{ga}. 

Other applications of the \xGm\ to realistic electronic structure calculations have recently appeared. 
B\"unemann, Weber and Gebhardt~\cite{Buneweber_ontheway, Bunemann,BuneWeb_LDAG_nonself1} 
implemented a non self-consistent Gutzwiller approach to electronic structure calculations, where a tight-binding 
model was set up from effective hopping parameters computed through a Kohn-Sham density functional calculation, 
and afterwards solved within the multi-band \ac{ga}.  
An approach where  both density and Gutzwiller parameters are optimized self-consistently was 
proposed in Refs. \onlinecite{Ho_LDAG_condmat} and \onlinecite{ZhongFang_LDAG}, and applied 
to several case studies.\cite{ZhongFang_LDAG_app1,Ho}
This method is in 
principle similar to ours, with the difference that it does not include the possibility of using a projector with 
nonzero off-diagonal matrix elements, which is instead a natural feature of our mixed-basis parametrization 
with $\hPhig$ operators. More recently, a fully unrestricted and density self-consistent Gutzwiller+LDA 
approach has been proposed\cite{Lanata_efficient} and applied to the $\gamma$-$\alpha$ iso-structural transition 
of Ce.\cite{Lanata-Kotliar} This method is in its formulation equivalent to ours, though the implementation is different. 

As an important and basic application,  
we applied the resulting \ac{ldag} method to the calculation of the electronic structure of $bcc$ Fe,
where important open questions about the role of correlations still remain, including the possibility of
an orbital-selective localization of $e_g$ electrons, see Ref. \onlinecite{anisimov} and references therein. 
Although we did not find an actual orbitally selective localization of electrons,  our results  confirm
that the magnetism of iron is, at least partially, driven by a double-exchange mechanism, caused by 
stronger localization of  $e_g$ states relative to $\tdg$ ones, a typically 
many-body phenomenon not described by conventional \ac{dft}. The double exchange mechanism  
would also arise as a direct consequence of a  selective $e_g$ Mott localization. 

The Gutzwiller approach enables to compute local magnetic moments, including their enhancement due to interactions, 
already at the unpolarized \ac{lda} level. The spin-polarized calculation separately provides the energy gain caused 
by interatomic magnetic alignment and ordering. The two phenomena, onset of magnetic moment and ordering, 
which 
come by necessity together and are treated on the same footing within simple \ac{lda}, \ac{lsda} and \ac{ldau}, 
are correctly very distinct within \ac{ldag}.
The present calculations of the electronic structure of Fe through \ac{ldag} implemented in the Siesta code 
could be further perfected. For example, and first of all, with the inclusion of two separate hopping renormalization factors 
on each $\eg$ and $\tdg$ multiplet of a double-$\zeta$ basis set, through which we will be able to better 
account for the effects of Hubbard-$U$ and exchange parameter $J$ on electron localization. The slightly excessive magnetic 
moment found can most likely be corrected through a better choice for the Hubbard-$U$ and 
by an improved evaluation of  
double-counting energy. In spite of the great number of parameters contained in $\hPhig$, the Lanczos-enhanced \ac{lm} algorithm 
we implemented here for the minimization of the energy with respect to Gutzwiller parameters is stable and fast, 
and can be easily parallelized to deal with more complex system as crystals having more than one atom per unit 
cell as transition metal compounds.

 \section{Acknowledgments}
We thank Nicola Lanat\`a for useful suggestions and discussions. 
This project was completed during the tenure  in Trieste of contracts PRIN/COFIN 2010LLKJBX 004, EU-Japan Project LEMSUPER, grant agreement 283214, 
Sinergia CRSII2136287/1 and ERC Advanced Grant 320796 MODPHYSFRICT.  
We also acknowledge financial support by the European Union, 
Seventh Framework Programme, under the project GO FAST, grant agreement no. 280555.
 
\appendix

\section{Imposing symmetries on the Gutzwiller projector}

The easiest basis in which to define the Gutzwiller projector is the basis of Slater determinants of single-particle wavefunctions, that we indicate in the text as \ac{bec}. A sample $N$-particle configuration on $d$ orbitals is for instance the $5$-electron, maximum-spin configuration, which in second quantized form reads
\begin{align}
\left(\prod_{m=-2}^2 \ocd_{m\up} \right) \ket{0}\,.
\end{align}
If we wish to write the most general projector in the \ac{bec}, the number of parameters we need is in principle equal to $2^{2(2l+1)}\times2^{2(2l+1)}$, where $2^{2(2l+1)}$ is the size of the many-body Hilbert space for multiplet of orbitals of angular momentum $l$. This is a huge number, which is hardly possible to treat in with numerical optimization already when $l=4$ or $5$.
In order to lower the amount of parameters that build up a Gutzwiller projector, we need to switch from this type of configuration basis, whose states are identified by single-particle quantum numbers as single particle spin $\sigma$, and magnetic quantum number $m$, to a basis of multi-particle quantum numbers that are good quantum numbers of the problem we are studying, which are, in the case of a system with full rotational invariance, the total spin $S^2$, the total angular momentum $L^2$, and the total spin and angular momentum magnetic quantum numbers $S_z$ and $L_z$. We will refer to this basis of many-body states labeled by good quantum numbers of the problem as to the \ac{mbsb}.
In the case of paramagnetic iron, the orbital rotational symmetry is broken by the cubic crystal field, resulting in a different set of conserved orbital quantum numbers, corresponding to the irreducible representations of the cubic group. In the case of spin-polarized iron, also the spin rotational invariance is broken in favor of a lower spin easy-axis symmetry, where only $S_z$ remains a good quantum number.
\subsection{Spin rotational symmetry}
It is well-known (see for instance Ref. \onlinecite{group_tableaux}) that the eigenstates of the total spin operator square $\Spinsq$ on the basis of a set of $N$ spins can be written in terms of Young tableaux. This is possible because of the isomorphism between the group $SU(N)$ and the irreducible representations of the permutation group, which are represented by Young tableaux. 
A general tableau provides a rule for the symmetrization-antisymmetrization with respect to particle exchange of a Slater determinant with a given number of electrons in a given orbital and spin configuration. Each box of a tableau corresponds to a particular filled single-particle orbital state, containing either a spin up ($\up$) or a spin down ($\dw$) electron. The orbitals belonging to the same row of a tableau must be symmetrized, while those belonging to different rows must be antisymmetrized. The many-body wavefunction produced by this symmetrization-antisymmetrization recipe turns out to be an eigenstate of both $\hat{S}^2$ and $\hat{S}_z$. The eigenvalue of $\hat{S}_z$ can be obtained by summing the spins in each box of the tableau, while the eigenvalue of $\hat{S}^2$ corresponds to the tableau shape.
For instance, the state with maximum $\tspinz$ component on $d$ orbitals is built from the totally symmetric tableau
\begin{align}
\young(\up\up\up\up\up)\,,
\end{align}
which corresponds to $S=5/2$. This particular state is a single Slater determinant, already belonging to basis of configurations.
The row-wise antisymmetrization rule of a tableau automatically imposes Pauli principle on the wavefunction, so that only wavefunctions obtained from one-column tableaux, or two-column tableaux with opposite spins on each column, are non-zero. For instance, the two-particle singlet state has a simple representation in terms of the totally antisymmetric two-particle tableau
\begin{align}
\young(\up,\dw)\,,
\end{align}
applied to a couple of electrons with opposite spin.
By application of raising and lowering operators $\hat{S}_{+}$ and $\hat{S}_{-}$ on a many-body wavefunction, one obtains another wavefunction which is generated by a tableau of the same shape. Every wavefunction with fixed values of $S$ and $S_z$ has an additional degeneracy which can be computed from the shape of the generating tableau, according to some simple rules~\cite{group_tableaux}.
\subsection{Implementation of crystal point symmetry}\label{Sec:orbrot_gutz}
In order to provide a classification of many-body wavefunctions according to point group quantum numbers, it is necessary to label them with angular momentum quantum numbers.
\subsubsection{Building eigenstates of angular momentum}
Thanks to Young tableaux we are able to label states with the quantum numbers $\{N, S^2,S_z, L_z \}$. For each of these sets of quantum numbers, there are several states with different values of the square modulus $L(L+1)$ of total angular momentum.
If the BC of our problem is already built from single-particle eigenstates of $\Elz$ and $\Elsq$, as for instance the $3d$ orbitals of a transition metal ($l=2$), it is very easy to build the angular momentum raising operator explicitly
\begin{align}
\Elpl &= \sum_{m=-l}^{l-1} \sqrt{l(l+1)-m(m+1)} \,\ocd_{m+1}\oc_{m}\,.
% \Elz &=\sum_{m=-l}^{l} m \ocd_{m}\oc_{m}\,.
\end{align}
From $\Elsq = \Elpl\Elmi + \Elz(\Elz-1)$ we can build the operator $\Elsq$, which will be block-diagonal in every subspace with fixed $\{N, S^2,S_z, L_z \}$. The diagonalization of every block gives the desired set of states, labeled by $\{N, S^2,S_z, L_z \}$. 
For large many-body spaces, as for instance the one built from $d$-electrons of a transition metal, another index $\theta$ might be needed, in order to distinguish between different states having the same set of quantum numbers listed above.
\subsection{Building eigenstates of point group symmetry operators}
% acronym MBSB
Provided that a set of many-body eigenstates of spin and angular momentum operators has been given, it is easy to break the rotational symmetry of the \ac{mbsb} in favor of some lower crystal symmetry when necessary. In this section we will treat the case of cubic symmetry, which is the case of iron. The ingredients we need for this purpose are just the following:
\begin{enumerate}
\item the $3\times 3$ matrix representation $G(g)_{ij}$ of the action of each element $g$ of the cubic group on a three-dimensional vector $\br$\,,
\item the character table of the group, for the cubic group it is shown in \rtab{Tab_cubic_chartab}\,,
\item the $\br$-space representation in spherical coordinates of an external potential with the symmetry of the group; an example for a potential with cubic symmetry is
\begin{align}
v[\hat{\br}(\theta,\phi)] = \cos(\theta)^4+\frac{1}{4}[3+\cos(4\phi)]\sin(\theta)^4\,,
\end{align}
where $\hat{\br}$ is the radial unit vector.
\end{enumerate}
\begin{table}[H]
\begin{center}
\begin{tabular}{|c|c|c|c|c|c|}
\hline
 &$E$& 8$C_3$ &3$C_2 (C^2_4)$ & 6$C_2$& 6$C_4$\\
\hline
$A_1$& 1& 1& 1& 1& 1\\ 
$A_2$&1& 1& 1& -1& -1\\
$E$&2& -1& 2& 0& 0\\
$T_1$& 3& 0& -1& -1& 1\\
$T_2$&3& 0& -1& 1& -1\\
\hline
\end{tabular}\caption{(Color online) Character table of the cubic group. The first row lists all the group classes along with the number of symmetry operations they contain. The following rows list the \irrep s, and their character on each symmetry class. From reference~\cite{Tinkham}.}\label{Tab_cubic_chartab}
\end{center}
\end{table}

Once these three ingredients are at hand, we proceed as follows:
\begin{itemize}
\item for each set of spherical harmonics $Y_{L,m}(\theta,\phi)$ with given $L$, we compute (by means of the algorithm of Gimbutas {\it et al.}~\cite{Gimbutas}) and diagonalize the matrix
\begin{align}
C^{(L)}_{m,m'} = \int  Y^\ast_{L,m}(\hat{\br}) v(\hat{\br}) Y_{L,m'}(\hat{\br})\,d\Omega\,;
\end{align}
\item for each set of spherical harmonics with given $l$ and for each group element $g$, we calculate the matrix elements
\begin{align}
M(g)^{L}_{m,m'} = \int  Y^\ast_{L,m}(\hat{\br}) Y_{L,m'}(G(g)^{-1} \hat{\br})\,d\Omega\,;
\end{align}
\item for each eigenvalue $\varepsilon$ of the matrix $C^{(L)}$, and for all eigenvectors $c^{\varepsilon,L,i}$ relative to this eigenvalue, we compute the character
\begin{align}
\chi({\cal C},L,\varepsilon) = \sum_{i}\sum_j c^{\varepsilon,L,i}_{j} M(g\in {\cal C})^{L}_{jk}c^{\varepsilon,L,i}_{k}
\end{align}
relative to the class ${\cal C}$. The value of the character enables us to assign the correct label of \irrep\ ${\cal I}$ to the eigenvectors $c^{\varepsilon,L,i}$.
\end{itemize}
The matrices $U^{(L)}_{ij}=c^{\varepsilon,L,i}_{j}$ are the unitary matrices we need to apply to every block of many-body basis states with a given value of $\telsq$ in order to switch from a basis labeled with $\{N, S^2, S_z, L^2, L_z\}$ to a basis indicated by $\{N, S^2,S_z, L^2, {\cal I}, \iota \}$\footnote{The quantum number $\telsq$ is still used to label states since each \irrep\ of the cubic group comes from a definite representation of the rotation group \Otre. However, in the case of cubic symmetry $\telsq$ is no longer a conserved quantum number, and the ground-state of the Hamiltonian will not necessarily have a definite $\telsq$.}, where $\iota$ labels the states within the same \irrep\ ${\cal I}$.
\section{Building the most general \xGp\ }\label{Sec_parametrization}
In this section we show how to parametrize the matrix $\hPhig$ of Gutzwiller parameters in the case of full spin and orbital rotational symmetry. The procedure is similar in the case of cubic symmetry.

We can easily construct the most general \xGp\ $\hPhig$ commuting with the operators $\Spinsq$, $\Elsq$, $\hat{S}_{x,y,z}$ and $\hat{L}_{x,y,z}$ by the following procedure. Operatively,
\begin{enumerate}
\item we find the quantum numbers that uniquely identify the \irrep\ of the symmetry group, in this case spin and spatial rotations \Sudu\ $\times$ \Otre\ . These quantum numbers are $\balpha= \{\Numbei, \tspinsq, \telsq \}$. The same representation can appear multiple times, so we will add another quantum number $\theta$ to distinguish between equivalent representations. 
Each \irrep\ has a degeneracy $n_{\{\balpha,\theta\}} = \telei\times\tspinei$; we will distinguish between states that are a basis for the same \irrep\ $\{\balpha,\theta\}= \{\Numbei, \tspinsq, \telsq, \theta \}$ through the index $\iota = \iota(\balpha\theta)$. In the case of spin and rotational symmetry $\iota$ lists all the eigenstates of $\Spinz$ and $\Elz$ within the same  $\tspinsq$ and $\telsq$.
\item With the previous definitions, the matrix elements of $\hPhig$ are labeled
\begin{align}\label{Eq_phi_sym}
\Phig_{\balpha\theta\iota,\bbeta\theta'\iota'} = \delta_{\balpha\bbeta}\delta_{\iota\iota'} \phi^{\balpha}_{\theta\theta'}\,,
\end{align}
where $\phi^{\balpha}_{\theta\theta'}$ is a reduced matrix element.
The labels $\balpha\theta\iota$ and $\bbeta\theta'\iota'$ identify univocally one state of the \ac{mbsb}, so that our parametrization of $\hPhig$ is complete.
\end{enumerate}

The same recipe holds when the spatial symmetry is, for example, the crystal cubic symmetry. In this case $\balpha= \{\Numbei, \tspinsq, {\cal I} \}$.

The result expressed by \eqn{Eq_phi_sym} comes directly from Schur's lemma, which states that a matrix commuting all the matrices of an \irrep\ of a group ${\cal G}$ must be a multiple of identity. 
The matrix $\Phig_{\balpha\theta\iota,\bbeta\theta'\iota'}$ must be nonzero only for $\balpha=\bbeta$ since, if $\hat{G}$ is a generator of the group and $\varepsilon_{\balpha}$ its eigenvalue with respect to any basis vector belonging to \irrep\ $\alpha$, the commutation relations $[\hPhig ,\hat{G}]=0$ imply that
\begin{align}\label{Eq_phi_sameeig}
\hat{G} \hPhig \ket{\balpha}  = \hPhig \hat{G} \ket{\Psi_{\balpha}} = \varepsilon_\balpha \hPhig \ket{\Psi_{\balpha}}\,
\end{align}
and that $\hPhig \ket{\balpha}$ must be a vector with the same quantum numbers $\balpha$. 

Again from the condition of zero commutator, we have that $\Phig_{\balpha\theta\iota,\balpha\theta'\iota'}$, seen as a matrix in the indices $\iota\iota'$ with fixed $\theta=\theta'$, must commute with all the matrices of \irrep\ $\balpha$, and by Schur's lemma it must be a multiple of the identity matrix. For $\theta\neq\theta'$ the same statement does not hold, since the representations are distinct. 

However, their equivalence implies that the matrices of the first are related to the matrices of the second through a unitary transformation. We can choose this transformation to be the identity, and this enables us to draw for $\theta\neq\theta'$ the same conclusions as for $\theta=\theta'$, so that $\Phig_{\balpha\theta\iota,\balpha\theta'\iota'}$ is diagonal in $\iota\iota'$ irrespectively of $\theta$ and $\theta'$.

\subsection{Symmetry reduction of parameter space}
The procedure explained in the previous paragraphs enables to considerably reduce 
the number of parameters for the Gutzwiller projector, so that its numerical optimization becomes not only computationally feasible, but also reasonably cheap.
In \rtab{Tab_parameter_number} we list the sizes of local many-body irreducible representations and the number of independent Gutzwiller parameters compatible with a few different spin and point symmetries.

\begin{table}[H]
\begin{center}
\begin{tabular}{|c|c|c|c|}
\hline
 spin symmetry & point symmetry & \# Hilbert & \# $\Phig$\\
\hline
SU(2)&O(3)& 81& 121\\
\hline
U(1)&O(3)& 176 & 336\\
\hline
SU(2)&cubic& 197& 873\\ 
\hline
U(1)&cubic& 428& 2716\\
\hline
% \hline
\end{tabular}\caption{Number of many-body irreducible representations generated by $d$ electrons and size of Gutzwiller parameter space for different types of spin (first column) and point (second column) symmetries. The symbol U(1) refers to axial spin symmetry, SU(2) to full rotational symmetry, O(3) to full spatial rotational symmetry.}\label{Tab_parameter_number}
\end{center}
\end{table}

\section{Sparse-constrained Levenberg Marquardt algorithm}\label{lm}

This algorithm performs the minimization of the Gutzwiller variational energy \eqn{Eq:Gutzfunc_LDA} with respect to the matrix elements of $\hPhig$. 
The details of the conventional constrained Levenberg-Marquardt (LM) algorithm are well explained by Fletcher~\cite{Fletcher}, who suggests the Multiplier Penalty Functional method (also known as Augmented Lagrangian method) as a way to impose constraints.
\subsection*{Levenberg-Marquardt algorithm with Lanczos approximation for the Hessian}
Depending on the quantity of single-particle orbitals involved in the definition of the \xGp, the number of parameters $x_i$ in the block-diagonal matrix $\Phig_{\alpha\beta}$ can be very large, which makes it computationally very expensive to compute the inverse Hessian matrix $\bm{h}^{-1}$ which is needed in the LM algorithm in order to find the descent direction $\bm{\delta}$, from the equation
\begin{align}\label{Eq:reduced_LMsystem}
\sum_j h_{ij} \delta_j = -g_{i}\,,
\end{align}
where $\bm{g}$ is the gradient of Gutzwiller Variational Energy with respect to Gutzwiller parameters.
Provided that $\bm{h}$ is positive-definite (and it can be modified to be so if necessary~\cite{Fletcher}), it can be convenient to solve \eqn{Eq:reduced_LMsystem} within a smaller parameter space, defined by taking several Lanczos steps through the Hessian matrix $\bm{h}$.
Also the memory storage of the algorithm can take great advantage of this possibility, since the definition of the Lanczos basis does not have as a requirement the knowledge of the full matrix $h_{ij}$, but only the knowledge of products $h_{ij}x_j$. 
Keeping in memory the full Hessian matrix is possible only for a small number of parameters, while it implies a considerable slow down of simulations in the case of a 5-band Gutzwiller projector like the one we need for dealing with transition metals.
Whenever we choose the starting Lanczos vector, we need to remember that finding an accurate solution for \eqn{Eq:reduced_LMsystem} requires the solution vector $\bm{\delta}$ to have a nonzero component on the first vector $\bm{x}$ of the Lanczos chain. It can be shown that, provided $\bm h$ is positive definite, the choice of the gradient $\bm{g}$ as starting vector ensures that  $\bm{\delta}$ has nonzero components on the first three vectors of the Lanczos chain.
Indeed, from the positive definiteness of $\bm h$ descends that
\begin{align}
\sum_{ij} \delta^\ast_i h_{ij} \delta_j > 0\,,
\end{align}
but since $\bm \delta$ must be such that $\sum_{j}h_{ij} \delta_j = -g_i$ (see \eqn{Eq:reduced_LMsystem}), we have that
\begin{align}
g^\ast_i \delta_i < 0\,,
\end{align}
so that $\bm g$ has a nonzero component on $\bm \delta$.
But we can say more than this, namely that there is a nonzero component of $\bm \delta$ also on $\bm{hg}$, since
\begin{align}
\sum_{ij}\delta^\ast_j h_{ij} g_i = -\sum_{ij} (g^\ast_i h_{ij} \delta_j)^\ast  = -\sum_j g_j g^\ast_j < 0\,
\end{align}
provided that the gradient is finite.
Finally, there is a nonzero component of $\bm \delta$ also on $\bm{h}^2 \bm{g}$, again due to the positive-definiteness of the Hessian, indeed
\begin{align}
\sum_{ij}\delta^\ast_i \left[h^2\right]_{ij} g_j &= \sum_{ij} \left\{g^\ast_i \left[h^2\right]_{ij} \delta_{j}\right\}^\ast  =\nonumber\\
&= -\sum_i \left\{g^\ast_i h_{ij} g_j\right\}^\ast < 0\,.
\end{align}
This means that three Lanczos steps will certainly improve a steepest descent problem. Any further step will further refine the approximation to the correct descent direction $\bm \delta$.
With the choice of the gradient as starting vector for the Lanczos chain, this minimization algorithm reduces to a constrained steepest descent in the limit of a single-vector Lanczos chain.
\vspace{0.2cm}

\acrodef{nsb}[NSB]{natural single-particle basis}
\acrodef{osb}[OSB]{original single-particle basis}
\acrodef{lm}[LM]{Levenberg-Marquardt}
\bibliographystyle{apsrev}
\bibliography{biblio}

\end{document}